\RequirePackage{fix-cm}
\documentclass[smallextended]{svjour3}       %
\smartqed  %

\usepackage{graphicx}
\usepackage{appendix}
\usepackage[round,sort]{natbib}
\setcitestyle{aysep={}} 
\usepackage{algorithm}
\usepackage{algorithmicx}
\usepackage{algpseudocode}

\usepackage[labelfont=bf]{caption}
\renewcommand\tablename{Table}
\usepackage{amsmath}
\usepackage{subcaption}
\usepackage{booktabs,caption}
\usepackage[flushleft]{threeparttable}
\usepackage{color}
\usepackage{url} 
\usepackage
[
        a4paper,%
        left=2.5cm,
        right=2.5cm,
        top=3cm,
        bottom=4cm,
]
{geometry}

\begin{document}

\title{A network approach to expertise retrieval based on path similarity and credit allocation%
}

\titlerunning{Expertise retrieval}        %

\author{Xiancheng Li \and Luca Verginer \and Massimo Riccaboni \and Pietro Panzarasa}%

\institute{Xiancheng Li \at
              School of Business and Management\\
              Queen Mary University of London, London \\
              \email{xiancheng.li@qmul.ac.uk}
          \and
          Luca Verginer \at
              Chair of Systems Design\\
              ETH Z\"urich, Z\"urich, Switzerland \\
          \and
          Massimo Riccaboni \at
              IMT School for Advanced Studies\\ 
              Lucca, Italy\\
          \and
          Pietro Panzarasa \at
              School of Business and Management \\
              Queen Mary University of London, London \\
 }

\maketitle

\begin{abstract}
With the increasing availability of online scholarly databases, publication records can be easily extracted and analysed. Researchers can promptly keep abreast of others' scientific production and, in principle, can select new collaborators and build new research teams. A critical factor one should consider when contemplating new potential collaborations is the possibility of unambiguously defining the expertise of other researchers. While some organisations have established database systems to enable their members to manually produce a profile, maintaining such systems is time-consuming and costly. Therefore, there has been a growing interest in retrieving expertise through automated approaches. Indeed, the identification of researchers' expertise is of great value in many applications, such as identifying qualified experts to supervise new researchers, assigning manuscripts to reviewers, and forming a qualified team. \textcolor{black}{Here}, we propose a network-based approach to the construction of authors' expertise profiles. Using the MEDLINE corpus as an example, we show that our method can be applied to a number of widely used data sets and outperforms other methods traditionally used for expertise identification.

\end{abstract}

\section{Introduction}

The increasing complexity of research problems calls for innovative solutions which combine knowledge from different scientific disciplines~\citep{van2011factors}. Therefore, many researchers become involved in interdisciplinary projects, thus collaborating with people with a variety of expertise. When facing the task of finding collaborators, scholars need to answer two inter-related questions: 1) How to identify an expert, i.e., how to find someone who is competent in a given field; and 2) how to profile an expert, i.e., how to identify the fields in which a given scholar is an expert. \textcolor{black}{In general, both questions jointly describe the objective of expertise retrieval~\citep{balog2012expertise}. Indeed figuring out the research area associated with} an individual represents a challenging research problem. Search engines such as Google Scholar or DBLP are of great help for finding documents~\citep{hertzum2000information}. However, these engines only return scientific documents, not the specific expertise of people. Even in an academic environment, researchers still have to rely on their social networks to identify the expertise of others~\citep{hofmann2010contextual}. 

\textcolor{black}{Identifying experts is crucial for academic groups when they need to involve a collaborator with specific expertise. In organisational settings, knowing the expertise of relevant researchers facilitates the assignment of important roles and jobs. For example, conference organisers may search for moderators, session chairs and keynote speakers with the proper expertise. And universities may want to recruit researchers with expertise in a particular fast-developing area to improve their reputation. A good method for expertise retrieval is therefore fundamental to provide the necessary knowledge for such activities.}  

However, expertise retrieval is challenging for many reasons. First, expertise is a relatively abstract concept, and there is currently no consensus on how to define it. Besides, expertise is a particular kind of knowledge stored in one's mind, and thus hard to \textcolor{black}{identify.} The only way to access people's expertise is through their works, e.g., documents, books, articles. Second, experts' names are often ambiguous. A single name may belong to multiple people, and the name of the same expert can vary in different databases. Indeed name disambiguation \textcolor{black}{has recently become a specific and independent} area of enquiry, and many studies have been carried out in this field~\citep{smalheiser2009author}. Finally, it is difficult to evaluate the strength of the association between an expert and the works he or she has been involved in, especially because an increasing amount of scientific production is co-authored by multiple individuals. Those challenges have made expertise retrieval a multi-faceted research area. In particular, since we learn about researchers' expertise mainly from their publications, the task of expertise retrieval has mainly been articulated into identifying the knowledge areas/topics in the text corpus and assigning them to the researchers~\citep{silva2018hierarchical}.

Inspired by previous approaches to dealing with credit allocation~\citep{shen2014collective} and by recent studies on finding node similarity in heterogeneous information networks (HIN)~\citep{shi2014hetesim}, we formalise the topics/expertise extracted from a given scientific publication as credit to be assigned to the co-authors of the publication, and propose a new method to allocate them to the co-authors based on their publication histories. Traditional approaches to the identification of the knowledge areas within the text corpus use topic-modelling methods such as Latent Dirichlet Allocation (LDA) based on controlled vocabulary from well-known classification systems such as the Medical Subject Headings ($MeSH$) in MEDLINE\footnote{https://www.nlm.nih.gov/MeSH/MeSHhome.html} and the topic tags in Microsoft Academic Graph (MAG)\footnote{https://academic.microsoft.com/topics}. 

Our work focuses on the process of evaluating the degree of each co-author's contribution to a collaborative work. We propose a new method for properly assigning the expertise to each co-author according to his or her contribution. Our method differs from traditional ones where the contribution of authors \textcolor{black}{is} assumed to be equal or assessed simply based on the order of authors in the byline. Moreover, our method can deal with large-scale data sets, and produces results that vary dynamically as the data set is updated over time. Unlike some citation-based approaches to the assessment of contributions, which require a certain time to account for the citations that accumulate over time, our method is experience-based and the update of authors' expertise is determined once the new records are added into the data set.

The rest of the article is organised as follows. In Section~\ref{sec:review} we review strengths and limitations of existing literature on expertise identification, and motivate our work. In Section~\ref{sec:data} we introduce the data used in our study. In Section~\ref{sec:algo} and Section~\ref{sec:sub} we present our new method and different %
\textcolor{black}{selection strategies.} In Section~\ref{sec:ext}, we provide some extensions to account for weights and time. In Section~\ref{sec:results} we report results obtained using the MEDLINE corpus and various examples. Section~\ref{sec:conclusion} summarises the findings of this work and outlines their implications for research and practice.

\section{Literature review}
\label{sec:review}

Previous work on expert profiling has primarily focused on identifying and ranking topics for a given expert~\citep{balog2007determining,serdyukov2011automatic}. However, only few studies have considered the temporal aspects of expertise. The work by~\cite{tsatsaronis2011become} was one of the first studies which focused on the evolution of authors' expertise over time. Their work was based on co-authorship information, and proposed evolution indices to measure the dynamics of authors' expertise. Inspired by their work,~\cite{rybak2014temporal} constructed temporal hierarchical expertise profiles using topic models. \textcolor{black}{Typically,} the underlying question of expert profiling is: What topics does a person know about?~\citep{balog2007determining,rybak2014temporal}. Indeed the word ``topic'' is commonly used in the various definitions of expertise because the traditional approaches to expertise profiling rely on topic models and Natural Language Processing (NLP) techniques~\citep{van2016unsupervised}. The main purpose of using those models is to classify documents into a number of topics and find a better match between authors and topics according to the topics extracted from their documents. As most of the machine learning algorithms belong to unsupervised learning, the topics are simply collections of words and thus not always appropriate for identifying expertise~\citep{silva2018hierarchical}. 

Since the main focus of expertise retrieval tasks is on the analysis of the documents, NLP techniques have commonly been applied. Traditional approaches to the expert profiling tasks are based on the LDA algorithm. LDA is a generative statistical model, first proposed in 2003, which considers each document as a mixture of a small number of topics and according to which the presence of each word is attributable to one of \textcolor{black}{the} topics of the document~\citep{blei2003latent}. LDA is a powerful tool to analyse documents and pinpoint topics, but it was not designed to address the task of identifying expertise. There is no better solution but to treat an author as a bigger document by combining all documents he or she has published. To include authorship information, ~\cite{rosen2004author} extended LDA and proposed the author-topic model for identifying the interests of authors. To make LDA suitable for different tasks in various contexts, many extensions have been proposed over the years. Some examples are the Author-Conference Topic model~\citep{tang2008topic}, the Author-Conference Topic-Connection model~\citep{wang2012author}, and the Author-Topic over Time model~\citep{xu2014author}. Some of these have been applied to practice as a part of a new search engine Aminer\footnote{https://aminer.org/}~\citep{tang2016aminer}.

However, classic LDA algorithms have several characteristics that are not ideal for such tasks. First, LDA requires a manual choice of the topic number. But one can hardly tell whether the choice is good or not since the performance of an LDA model is evaluated by \textit{perplexity}, a metric proposed by~\cite{blei2003latent}. Therefore it is difficult to decide and evaluate the number of topics. When such number is too large or too small, the research areas (corresponding to the topics) provided by LDA may become too general or too specific~\citep{berendsen2013assessment}. Second, since LDA is an unsupervised learning algorithm, topics generated from LDA are just distributions of words without labels which can be hard to interpret. Additionally, the academic research areas are always connected and have a hierarchical structure. However, LDA generates independent topics without any kind of relationships between them~\citep{silva2018hierarchical}. 

While most studies are concerned with better solutions to address the flaws of topic models, few have highlighted the importance of author-document connections in the tasks of expertise retrieval. In 2012,~\cite{duan2012mei} first integrated community discovery with topic modelling, and proposed the Mutual Enhanced Infinite Community-Topic model which finds communities and the topics they discuss in text-augmented social networks. Lately, more studies have started using information networks to avoid the problems of the LDA models. \cite{gerlach2018network} represent the data as a bipartite network of words and documents and convert the task into finding communities in such a network. Some different approaches that focus on topic modelling using HINs have been proposed~\citep{sun2009ranking}. Subsequently, a pioneer algorithm called Rankclus was designed. It uses a generative model that operates on bipartite topologies and simultaneous clusters and ranks nodes in a HIN~\citep{sun2009rankclus}. More recently, different community detection methods, such as generative model and modularity optimisation, have been applied to the creation of hierarchical expert profiles~\citep{wang2015constructing,silva2018hierarchical}.

Despite the efforts of many scholars to find better ways for extracting individuals' interests from the works they produced, most studies have paid little attention to the unequal contributions of authors in collaborative works. Authors that publish with other co-authors in several fields can be associated with multiple topics found in their publications. Identifying the expert on a specific field associated with a paper requires the identification of the different contributions of authors in collaborative works, and therefore identifying one or more people as experts bears \textcolor{black}{a} resemblance to a credit allocation problem. 

In the last decade, as the complexity and interdisciplinarity of modern research have \textcolor{black}{steadily risen}, collaborations among researchers have been playing an increasingly important role~\citep{newman2004coauthorship}. The multidisciplinary nature of research requires expertise from different scientific fields ~\citep{lawrence2007mismeasurement}. \textcolor{black}{In turn, as a result of the increasing size of the newly formed scientific groups, the scientific credit system has come under mounting pressure~\citep{koopman2010give}.} As a matter of fact, \textcolor{black}{the interdisciplinarity of modern science} not only endangers the current credit allocation system, but also poses more obstacles to expertise retrieval. In such interdisciplinary collaborations, authors from different fields work together to produce one result (e.g., an article), but each \textcolor{black}{author} contributes only partly to the publication. It can therefore be difficult to quantitatively discern the individual co-authors’ contributions to a multi-authored publication~\citep{bao2017dynamic}. Most topic models 
for expertise retrieval cannot solve this problem, and new approaches \textcolor{black}{to allocating scientific credit to co-authors} are therefore required.

Current approaches to credit allocation fall in several major categories. The first and classic one is to view each author as the sole author contributing a copy of the same publication. The second is to distribute the contribution to all co-authors evenly, and the third according to the order in the publication byline or to the role of the co-authors~\citep{hirsch2005index,hirsch2007does,stallings2013determining}. The first two categories are obviously biased to some degree, and the third is based on some acquiescent agreements according to disciplines which may not be easily acceptable by others. Recently, scholars have been working on allocating credit based on the specific contribution of each author~\citep{foulkes1996redefining,tscharntke2007author}.~\cite{shen2014collective} proposed a new method which focuses on the co-citations. This method is based on the intuition that the more an author appears in a co-cited paper, the more credit he or she should receive. And they managed to capture the contribution of co-authors as perceived by the scientific community and successfully tested on the Nobel Prize publications. Considering that the novelty of a paper and the attention paid to it tend to fade with time, \cite{bao2017dynamic} extended their idea and proposed a dynamic credit allocation algorithm.

As science can be regarded as a complex, self-organising and evolving network of scholars, projects, papers and ideas~\citep{fortunato2018science}, another way to deal with the unequal contributions of multiple authors to collaborative works is to use the similarity between a node representing a given topic and a node representing a given author to assess the contribution that the author made to the focal document with respect to the topic. Information networks are networks consisting of data items linked in some way. The best known example is the World Wide Web where the nodes are web pages consisting of texts, pictures or other information, and the links are hyperlinks that allow us to navigate from one page to another. There are some networks which could be considered information networks and also have social connotations. Examples include the networks of email communication, and online social networks such as Twitter and Facebook~\citep{xiong2015top}. 

\begin{figure}
    \centering
    \begin{subfigure}{.6\textwidth}
        \includegraphics[width=1\textwidth]{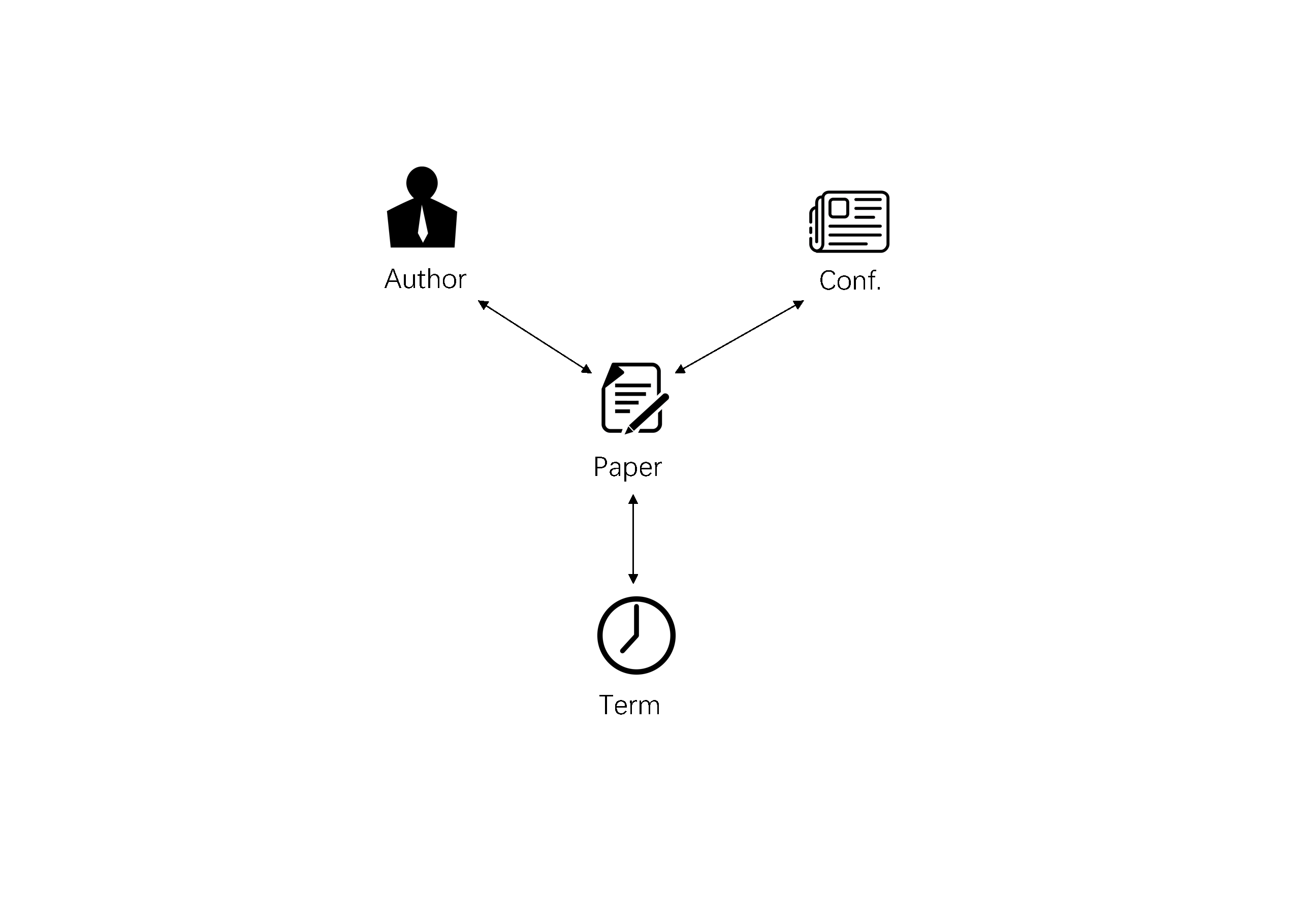}
        \caption{DBLP data}
        \label{subfig:DBLP}
    \end{subfigure}
    
    \begin{subfigure}{.6\textwidth}
        \includegraphics[width=1\textwidth]{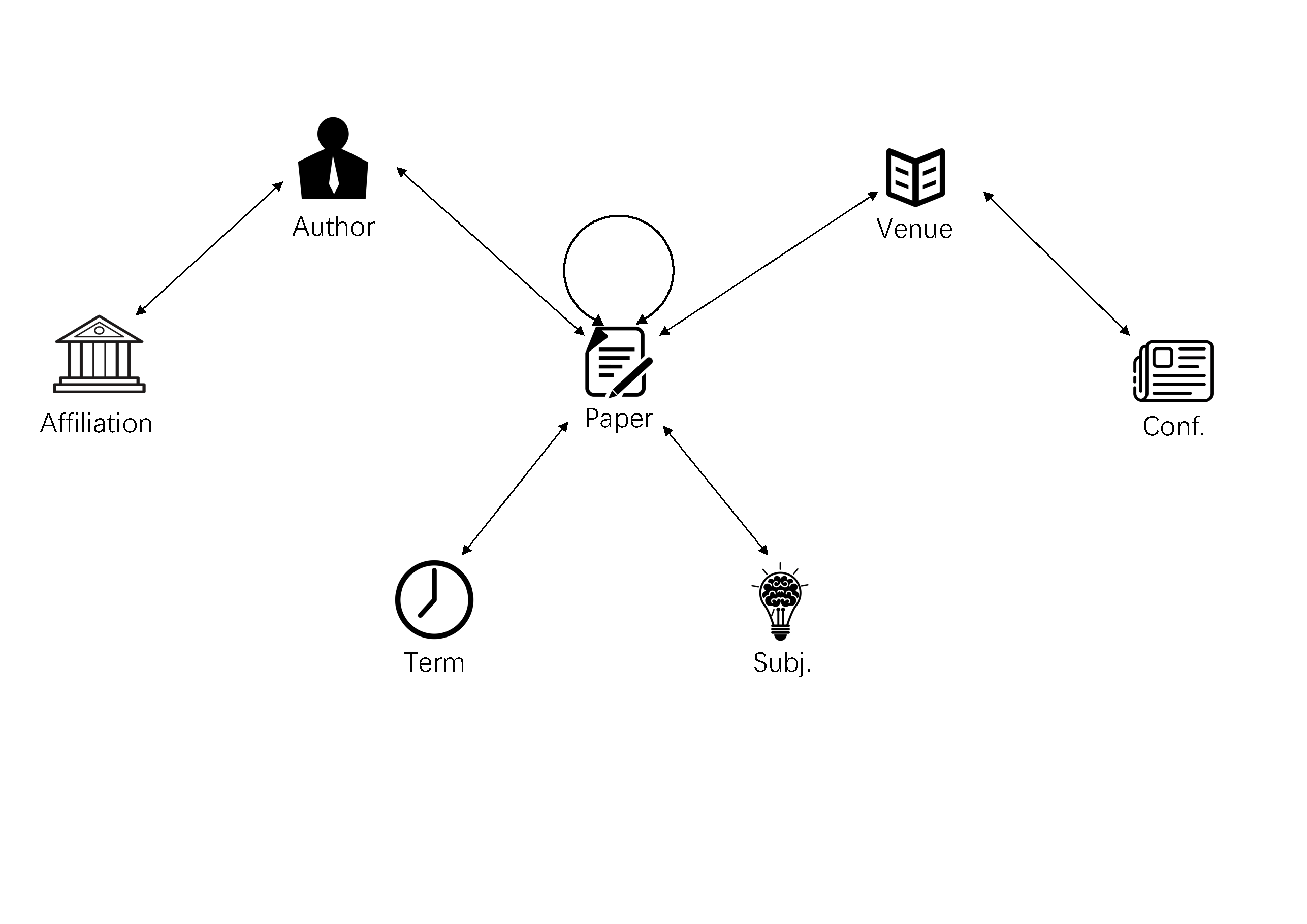}
        \caption{ACM data}
    \end{subfigure}
        \caption{Examples of typical Heterogeneous Information Networks (HINs)}
    \label{fig:ex_HINs}
    
\end{figure}

An information network is defined as a directed graph $G = (V, E)$ with an object type mapping function $\phi: V \rightarrow A $ and a link type mapping function $\psi(e): E \rightarrow R$, where each object $v \in V$ belongs to one particular object type $\phi(v) \in A$, and each link $e \in E$ belongs to a particular relation $\psi(e) \in R $. Unlike the traditional network definition, we explicitly distinguish object types and relationship types in the network. Notice that, if \textcolor{black}{there exists} a relation from type A to type B, denoted as $A\xrightarrow[]{R}B$, the inverse relation $R^{-1}$ holds naturally for $B\xrightarrow[]{R^{-1}}A$. Most of the time, $R$ and its inverse $R^{-1}$ are not equal, unless the two types are the same and $R$ is symmetric. When the types of objects $|A| > 1$ or the types of relations $|R| > 1$, the network is called heterogeneous information network (HIN); otherwise, it is a homogeneous information network. In real-world networks, multiple-typed objects are often interconnected, forming HINs~\citep{shi2012relevance}.
A bibliographic information network is a typical HIN, containing objects from several types of entities. The most common entities are papers ($P$), venues (conferences/journals) ($V$), authors ($A$), affiliations ($aff$), and terms ($T$). The DBLP and ACM data in Fig.~\ref{fig:ex_HINs} is a typical example~\citep{shi2014hetesim}. There are links connecting different-typed objects and the link types are defined by the relations between two object types. For a bibliographic network, links can exist between nodes of the same or different types. For example, there are links between authors and papers denoting the ``write'' or ``written-by'' relations, and links between papers denoting ``cite'' and ``cited-by'' relations. 

In a heterogeneous network, two objects can be connected via different paths. For example, two authors can be connected via the ``author-paper-author'' path, the ``author-paper-venue-paper-author'' path, and so forth. Formally, these paths are called \textit{meta-paths}. In a graph $TG = (A, R)$, where $A$ is the set of node types and $R$ is the set of relation types, a meta path $P$ is a path denoted in the form of $A_1\overset{R_1}{\rightarrow}A_2\overset{R_2}{\rightarrow}...\overset{R_l}{\rightarrow}A_{l+1}$, which defines a composite relation $R = R_1 \circ R_2 \circ...\circ R_l $ between type $A_1$ and $A_{l+1}$, where $\circ$ denotes the composition operator on relations~\textcolor{black}{\citep{shi2014hetesim}.}

Similarity search is a primitive operation in large-scale HINs that consist of multi-typed, interconnected objects, such as the bibliographic networks and social media networks. Traditional similarity measures (e.g., cosine similarity) are computed between vector representations of features, using numerical data types~\citep{nguyen2010cosine}. In information networks, however, the interconnections between objects are sometimes more important than the features of the objects themselves. 

To capture the information contained in the links, \cite{lin2006pagesim} proposed a link-based similarity measure \textit{PageSim} and applied it to the identification of similar web pages. PageSim only works on networks with one type of nodes (e.g., homogeneous information networks), but many networks are heterogeneous. Considering the semantics in meta paths constituted by different-typed objects, \cite{sun2011pathsim} first proposed the path-based similarity measure \textit{PathSim} to evaluate the similarity of same-typed objects based on symmetric paths. Following their work, \cite{yao2014pathsimext} extended PathSim by incorporating richer information, such as transitive similarity, temporal dynamics, and supportive attributes. A path-based similarity join method \textit{JoinSim} was proposed to return the top $k$-similar pairs of objects based on user-specified join paths~\citep{begummeta}. \cite{wang2016relsim} defined a meta-path-based relation similarity measure, \textit{RelSim}, to examine the similarity between relation instances in schema-rich HINs. In order to evaluate the relevance of different-typed objects, \cite{shi2014hetesim} proposed \textit{HeteSim} to measure the relevance of any object pair under arbitrary meta path. \textcolor{black}{To overcome the problem related to the high computational and memory requirements of \textit{HeteSim},}~\cite{meng2014relevance} proposed the \textit{AvgSim} measure that evaluates the similarity scores, respectively, through two random walk processes along the given meta path and the reverse meta path.

\textcolor{black}{The idea of node similarity can be useful in expertise retrieval because, if we can measure the similarity between a given author and a field, we can assess the author's expertise in that field. \textit{HeteSim} has been designed to evaluate the relevance of different-typed objects, and thus has the potential to be applied to the task of expertise retrieval. However, this task needs to explicitly account for the uneven contribution of various authors to collaborative efforts, and therefore cannot be carried out merely by applying simple measures of similarity between nodes. For this reason, we decided to draw on \textit{HeteSim}, and propose a properly adjusted method for capturing authors' expertise in evolving networks.}

As a result of the increasing interest in extracting relevant topics from scientific publications, many widely used online data sets provide external controlled vocabulary to classify publications. Some examples are the $MeSH$ classification system in MEDLINE and the topic tags in MAG. Those systems have used a variety of techniques to improve the reliability of the classifications, and some scholars have started to use them as ground truth or baseline in their works~\citep{alshebli2018preeminence}. Our method simplifies the process of topic extraction from documents by using the MEDLINE corpus as an example, \textcolor{black}{and focuses on how to allocate expertise to co-authors that unevenly contribute to collaborative efforts.}

\textcolor{black} {The method for collective credit allocation in science developed by~\citep{shen2014collective} is conceptually similar to our method. Yet, it differs from ours in one important aspect: 
it focuses on the process of appropriately allocating the credit of a given paper to each of the co-authors. It uses the co-citations to the given paper and other papers published by the co-authors to determine the proportion to be assigned to each co-author of the paper. If more papers have cited at the same time the focal paper and other papers published by a given co-author, a larger proportion of the credit will be allocated to this co-author, indicating a larger contribution is made by the co-author in this work. However, at the time when a paper is published and therefore has no citations, contributions to this paper are equally allocated across co-authors. Moreover, because the citations vary over the years, so does the credit allocated to each co-author by this method. Clearly, one shortcoming of this method lies on the fact that the contribution of an author to a paper should be unambiguously defined once the paper is published, and should therefore be assessed according to the experience or background of each co-author rather than based on future citations.}

\section{Data\label{sec:data}}

MEDLINE (Medical Literature Analysis and Retrieval System Online) is a bibliographic database of life sciences and biomedical information, \textcolor{black}{maintained and curated by the US National Library of Medicine.} It includes bibliographic information on articles from academic journals covering medicine, nursing, pharmacy, dentistry, veterinary medicine, and healthcare. The database contains records from more than $5,000$ selected journals covering biomedicine and health from 1948 to the present. The database is freely accessible via the PubMed interface\footnote{https://www.ncbi.nlm.nih.gov/pubmed}. 

In addition, PubMed provides an online scientific publication search engine that associates each paper with several $MeSH$ terms. These terms are similar to keywords of papers, except that a controlled vocabulary is used to classify publications. Since the $MeSH$ terms of a paper are not given by the authors, they are not subject to subjective biases and can be considered as labels which indicate the major topics discussed in the paper. PubMed also constructed tree structures for $MeSH$ terms\footnote{https://MeSHb.nlm.nih.gov/treeView} so that one can look for the research field of each $MeSH$ term.

\textcolor{black}{In particular, in PubMed, each $MeSH$ term has one $MeSH$ Unique ID (starting with letter `D' followed by $6$ digits) and at least one $MeSH$ Tree ID (starting with a letter followed by digits separated by dots). For example, the $MeSH$ Tree ID of `Anatomic Landmarks' is `A01.111' and its $MeSH$ Unique ID is `D059925'. The first letter of the $MeSH$ Tree ID of a $MeSH$ term indicates which one of the $16$ categories the $MeSH$ term belongs to.}\footnote{\textcolor{black}{The following are the $16$ most general categories: A. Anatomy; B. Organisms; C. Diseases; D. Chemicals and Drugs; E. Analytical, Diagnostic and Therapeutic Techniques and Equipment; F. Psychiatry and Psychology; G. Phenomena and Processes; H. Disciplines and Occupations; I. Anthropology, Education, Sociology and Social Phenomena; J. Technology, Industry, Agriculture; K. Humanities; L. Information Science; M. Named Groups; N. Health Care; V. Publication Characteristics; Z. Geographicals.}} 
However, the $MeSH$ terms in the raw data are indexed by the $MeSH$ \textcolor{black}{Unique} ID rather than the \textcolor{black}{$MeSH$} Tree ID. To map each $MeSH$ \textcolor{black}{Unique} ID with the corresponding \textcolor{black}{$MeSH$} Tree ID, we downloaded detailed information about each $MeSH$ \textcolor{black}{Unique} ID and used Regular Expression (Regex) to search the match between each $MeSH$ \textcolor{black}{Unique} ID and the corresponding \textcolor{black}{$MeSH$} Tree ID.%
\footnote{In cases where the $MeSH$ Unique ID has two $MeSH$ Tree IDs, we kept both $MeSH$ Tree IDs.} The $MeSH$ Tree ID can have a different depth (the depth of a node is the number of edges from the node to the tree's root node). Some $MeSH$ IDs have corresponding \textcolor{black}{$MeSH$} Tree IDs of depth five (e.g., 'A15.378.316.378'), others only have depth of two (e.g., 'B02'). To ensure that all $MeSH$ IDs can be mapped to the same depth of \textcolor{black}{$MeSH$} Tree IDs, we converted all $MeSH$ Tree IDs to depth two by cutting the numbers after the first point. As a result, all $MeSH$ IDs have been mapped to $127$ \textcolor{black}{$MeSH$} Tree IDs of depth two.

To disambiguate authors' names we used the data set provided by Torvik named Author-ity~\citep{torvik2009author}. The data set provides the disambiguated authors' names appearing in the MEDLINE data set up to the year 2008. In our work, we used the first decade of publications in MEDLINE, from 1948 to 1957, to test the method we developed and make a comparison between a baseline (\textit{BL}) method and our method. 

\section{\textit{HeteAlloc}: An algorithm based on path similarity\label{sec:algo}}

\subsection{The method}
Based on the idea described above, the task of expertise profiling can be transformed into a dynamic $MeSH$ terms allocation problem: given a time $T$, an author $A$ and a $MeSH$ term $M$, what is the expertise of author $A$ on $MeSH$ term $M$ at time $T$? To answer this question, we have developed a method based on the idea of credit allocation, using the author-paper and paper-$MeSH$ connections. %
Notice that what we care about is the effort devoted by an author to a \textcolor{black}{$MeSH$ term} (measured by the number of papers published with that \textcolor{black}{$MeSH$ term}, or possibly by the reputation or impact factor of the journals, research venues and outlets where these papers have appeared), rather than the reputation of the author (measured by the citations received). 

\begin{figure}[htp]
\centering
\includegraphics[width=0.6\textwidth]{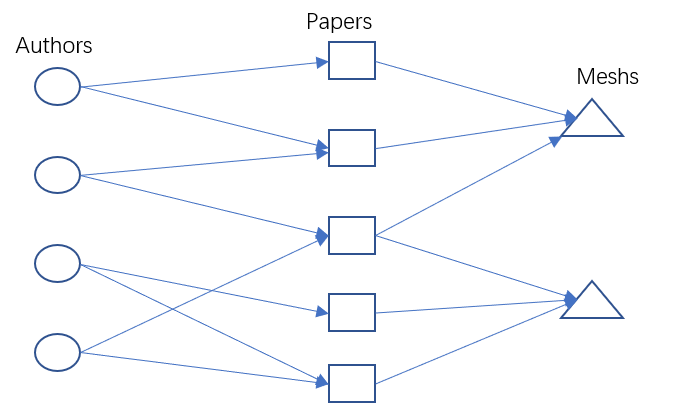}
\caption{An example of HIN}
\label{fig:HIN_demo}
\end{figure}

\textbf{Problem description.}
We focus on a subset of the HIN which contains three types of nodes: Papers, Authors and $MeSH$ terms. A simple example of this HIN is shown in Fig.~\ref{fig:HIN_demo}. In this network, the $MeSH$ terms are indexed by $MeSH$ tree IDs, and the links between papers and $MeSH$ terms show which $MeSH$ term the papers are associated with. Our problem is how to allocate credit to single authors. The input to this question is the link lists of every year between 1948 to 1957, and the output is a vector for each author with a value for each of the $127$ $MeSH$ categories indicating the author's expertise in those $MeSH$ categories. 

We developed a dynamic credit allocation algorithm based on Path Similarity which we shall call \textit{HeteAlloc}. Based on the HIN with three types of nodes (i.e., authors, papers and $MeSH$ terms), our task is to assign the credit of each $MeSH$ term in a paper to the corresponding authors, and to use the whole publication history of authors to find their expertise. Our method will calculate the similarity between an author and a $MeSH$ term, and assign a value to each author based on the similarity. It is based on \textit{HeteSim}~\citep{shi2014hetesim} as this method is able to measure the similarity between different types of nodes, i.e., authors and $MeSH$ terms in this case.

\textbf{Heterogeneous Similarity (\textit{HeteSim}).} \textit{HeteSim} is a measurement of the relatedness of heterogeneous objects based on an arbitrary search path. The properties of \textit{HeteSim} (e.g., symmetric and self-maximum) make it suitable for a number of applications. We define \textit{HeteSim} as follows:

\textbf{HeteSim}: Given a relevance path $P = R_1 \circ R_2 \circ\cdots R_l$, the \textit{HeteSim} score between two objects $s$ and $t$ ($s\in R_{1}.S$ and $t\in R_{l}.T$) is

\begin{equation}
\begin{split}
&HS(s,t|R_1 \circ R_2 \circ\cdots R_l) =\\
&    \frac{1}{\left | O(s|R_1)) \right |\left | I(t|R_l)) \right |}\sum_{i = 1}^{O(s|R_1)}\sum_{j=1}^{I(t|R_l)} HS(O_i(s|R_1),I_j(t|R_l)|R_2 \circ\cdots R_{l-1}),
\end{split}
    \label{SH}
\end{equation}

\noindent where $O(s|R_1$) is the out-neighbours of $s$ based on relation $R_1$, and $I(t|R_l)$ is the in-neighbours of $t$ based on relation $R_l$. 

\textcolor{black}{\textbf{Transition probability matrix}. The adjacent matrix $\textbf{W}_{AB}$ is defined for all links from nodes of type A to nodes of type B. The transition probability matrix $\textbf{U}_{AB}$ is the normalised matrix of $\textbf{W}_{AB}$ along the row vectors.}

\textcolor{black}{\textbf{Reachable probability matrix}. Given a network $G = (V, E)$ following a network schema $S = (A,R)$, a reachable probability matrix $PM$ for a path $P = (A1A2$ · · · $Al+1)$ is defined as $\textbf{PM}_P$= $\textbf{U}_{A1A2}$
$\textbf{U}_{A2A3}$ · · · $\textbf{U}_{A_lA_{l+1}}$. $\textbf{PM}(i, j)$ represents the probability of object $i \in A_1$ of reaching object $j \in A_{l+1}$ under the path $P$.}

Using the reachable probability matrices~\citep{ramage2009random}, the \textit{HeteSim} between two nodes $a$ and $b$ can be written in a matrix form as 

\begin{equation}
    \textcolor{black}{HeteSim(a,b|P) = \textbf{PM}_{P_L}(a,:)\textbf{PM}'_{P_{R^{-1}}}(b,:)},
\end{equation}
\noindent where $PM$ is the reachable probability matrix, and $PM_P(a,:)$ refers to the $a$-th row in $PM_P$.

Finally, Equation~\ref{Nor_SH} provides the normalised version of \textit{HeteSim}, which ensures that the similarity between a node and itself is equal to one
\begin{equation}
    \textcolor{black}{HeteSim(a,b|P) = \frac{\textbf{PM}_{P_L}(a,:)\textbf{PM}'_{P_{R^{-1}}}(b,:)}{\sqrt{\left \| \textbf{PM}'_{P_{R^{-1}}}(b,:) \right \|\left \| \textbf{PM}_{P_L}(a,:) \right \|}}}
    \label{Nor_SH}
\end{equation}
\newline

\textbf{\textit{HeteSim} in $MeSH$ term assignment.} The definition of \textit{HeteSim} in Equation~\ref{Nor_SH} can be directly applied to our network. For a node of type author ($A$) $a_0$ and a node of type $MeSH$ ($M$) $m_0$, the \textit{HeteSim} between $a_0$ and $m_0$ is

\begin{equation}
   \textcolor{black}{HeteSim(a_0,m_0 | a_0\in A,m_0 \in M) =  \frac{\textbf{M}_{AP}[a_0,:]\cdot \textbf{M}'_{MP}[m_0,:]}{\sqrt{\left \| \textbf{M}_{AP}[a_0,:] \right \|}\cdot \sqrt{\left \| \textbf{M}'_{MP}[m_0,:] \right \|}} },
    \label{HS1}
\end{equation}
\noindent where \textcolor{black}{$\textbf{M}_{AP}$ and $\textbf{M}_{MP}$} are adjacency matrices between the Author nodes, Paper nodes and between $MeSH$ nodes and Paper nodes, respectively. In Equation~\ref{HS1}, the adjacency matrix is used instead of the reachable probability matrix to make our method more interpretable. It can be shown that the formalisation of \textit{HeteSim} using the adjacency matrix can be the same in an unweighted network as the formalisation of \textit{HeteSim} based on the reachable probability matrix. Note that \textcolor{black}{$\textbf{M}_{MP}=\textbf{M}_{PM}^{'}$}, the matrix product resulting by multiplying \textcolor{black}{$\textbf{M}_{AP}$ and $\textbf{M}_{PM}^{'}$}, is the weighted reachable matrix between node type Author and node type $MeSH$. Formally, we have

\begin{equation}
    \textcolor{black}{N_{papers\ published\ by\ a_0\ which\ include\ m_0}=\ \textbf{M}_{AP}[a_0,:]\cdot \textbf{M}'_{MP}[m_0,:]},
\end{equation}
\noindent where $N$ means `the number of'.

Note that all elements in $\textbf{M}_{MP}$ and $\textbf{M}_{AP}$ are either $1$ or $0$, and thus we have
\begin{equation}
    \textcolor{black}{\left \| \textbf{M}_{AP}[a_0,:] \right \| = \sum \textbf{M}_{AP}[a_0,:]}.
\end{equation}

Thus,
\begin{equation}
    \textcolor{black}{\sqrt{\left \|  \textbf{M}_{AP}[a_0,:] \right \|}=\ \sqrt{\sum{\textbf{M}_{AP}\left[a_0,:\right]}}=\sqrt{N_{paper\ published\ by\ author\ a_0}}}.
\end{equation}

In the same way,
\begin{equation}
    \textcolor{black}{\sqrt{\left \|  \textbf{M}'_{MP}[a_0,:] \right \|}=\ \sqrt{\sum{\textbf{M}'_{MP}\left[a_0,:\right]}}=\sqrt{N_{paper\ which\ include\ the \ MeSH\ term\ m_0}}}.
\end{equation}

Equation~\ref{HS1} can therefore be rewritten as

\begin{equation}
    \textcolor{black}{HeteSim(a_0,m_0 | a_0\in A,m_0 \in M) =  \frac{\textbf{M}_{AP}[a_0,:]\cdot \textbf{M}'_{MP}[m_0,:]}{\sqrt{\sum \textbf{M}_{AP}[a_0,:]}\cdot \sqrt{ \sum \textbf{M}'_{MP}[m_0,:]}}},
    \label{HS2}
\end{equation}

and interpreted as

\begin{equation}
\begin{split}
&    HeteSim(a_0,m_0| a_0\in A,m_0 \in M) = \\ & \frac{N_{\textup{papers published by author}a_0 \textup{ which include the $MeSH$ 
 }m_0}}{\sqrt{N_{\textup{papers published by author }a_0}}\cdot \sqrt{N_{\textup{papers which include the $MeSH$ term 
 }m_0}}}. 
 \end{split}
    \label{inter_HS}
\end{equation}

Though \textit{HeteSim} is quite suitable for our task, there are some disadvantages. The most important one is that \textit{HeteSim} is a ``global'' measure in a sense. When the similarity between an author and a $MeSH$ term is calculated, all papers are taken into consideration, even those which have no connection with the target author. For example, if someone published a paper with a $MeSH$ term $M1$, the similarity of all authors with $M1$ will decrease even if none of them has ever worked with him or her. As a matter of fact, the original \textit{HeteSim} measures the contribution of each author to the total knowledge (limited in the data set) of a $MeSH$ term. However, the expertise we want to examine refers to the \textcolor{black}{$MeSH$ term} where an author conducted most of his or her work. In a real-world situation, one can only contribute to several hundreds of papers at most. And if we compare this fraction of papers to the tremendous overall amount of papers available in online databases, the similarity will be significantly small and the original \textit{HeteSim} will have a poor performance. \\

\textbf{Modification of \textit{HeteSim} (\textit{HeteAlloc}).} To address this shortcoming of \textit{HeteSim}, here we propose a modified version, namely \textit{HeteAlloc}. The underlying idea is to limit the calculation to a subset of papers, which can be selected according to the context. Formally, we have

\begin{equation}
\begin{split}
    &{\rm{HeteAlloc(a}},{\rm{\;m }}\left| {{\rm{ a\;}} \in {\rm{A}},{\rm{\;m\;}} \in {\rm{M}}} \right.{\rm{ }})= \\
    &\frac{{\textbf{M}_{AP}}\left[ {a,:} \right] \cdot ({\textbf{M}_{sub}}\left[ {a,:} \right] \odot \textbf{M}_{MP}\left[ {m,:} \right])'}{{\sqrt  {\left \|  {\textbf{M}_{AP}}\left[ {a,:} \right] \right \|} \; \cdot \;\sqrt {\left \| {\textbf{M}_{sub}}\left[ {a,:} \right] \odot {\textbf{M}_{MP}}\left[ {m,:} \right] \right \|} }},
    \end{split}
    \label{General_SHS}
\end{equation}
\newline
\noindent where the operation $\odot$ is the element-wise product, and $M_{sub}$ is the subset selection matrix with

\begin{equation}
\begin{split}
   & \textbf{M}_{sub}\left[a,n\right]=\\
   & \begin{cases}
1 & \text{if the $n^{th}$ paper is in the selected subset of target author a}   \\ 
0 & \text{ otherwise }  
\end{cases} 
\end{split}
\end{equation}

Like the original \textit{HeteSim}, our method is based on the cosine of two vectors. As \cite{pirotte2007random} pointed out, the angle between the node vectors is a much more predictive measure than the distance between the nodes. The only difference is that the second vector is filtered by a row of subset selection matrix. The selection of the subset is the essential part of our method, and requires a considerable amount of effort towards the design and computation of the matrix multiplication. 

In what follows, we shall present three subset selection strategies, and then show how to compute the measure, discuss the advantages and disadvantages of each strategy, and finally provide interpretations.

\section{Subset selection strategies}\label{sec:sub}
\subsection{Subset of co-authors' papers.} The basic idea of this strategy is that only those who have co-authored with the focal author should be entitled to influence the assignment of his or her expertise. The \textit{HeteSim} measure should therefore be limited to the subset of papers published either by our target author or those who have co-authored with this author. To find the subset, we provide the following definition:
    
    \textbf{Binary Reachable Matrix of Path Length $i$}: Given relation $A \overset{R}\rightarrow B$ and the adjacency matrix $\textbf{W}_{AB}$ between type $A$ and type $B$, the Binary Reachable Matrix of Path Length $i$ from $A$ to $B$ following meta-path $AB^{i}$ is

    \begin{equation}
        \textbf{RM}_{AB}^{(i)}\left(m,n\right)= 
    \begin{cases}    
    0 & \text{ if } \textbf{M}^{(i)}_{AB}(m,n)=0  \\ 
    1 & \text{otherwise }  
    \end{cases} 
    \end{equation}
    
    \noindent where $\textbf{M}^{(i)}_{AB}= \textbf{W}_{AB}\cdot (\textbf{W}_{BA}\cdot \textbf{W}_{AB})^{(i-1)}$.

The selected subset, ${\textbf{RM}}_{AP}^{2}$, follows the meta-path `APAP', which, for each author, creates the subset of papers published by the author or his/her co-authors. To be more specific, the $n$-th row of ${\textbf{RM}}_{AP}^{2}$ is a vector where the $m^{th}$ value is $1$ if, for the $n$-th author, paper $m$ is included in the subset. To this end, we define \textit{HeteAlloc}
    
    \begin{equation}
    \begin{split}
    &{\rm{HeteAlloc(a}},{\rm{\;m }}\left| {{\rm{ a\;}} \in {\rm{A}},{\rm{\;m\;}} \in {\rm{M}}} \right.{\rm{ }}) =\\
    &\frac{{\textbf{M}_{AP}}\left[ {a,:} \right] \cdot ({\textbf{RM}_{AP}^{(2)}}\left[ {a,:} \right] \odot \textbf{M}_{MP}\left[ {m,:} \right])'}{{\sqrt  {\left \|  {\textbf{M}_{AP}}\left[ {a,:} \right] \right \|} \; \cdot \;\sqrt {\left \| {\textbf{RM}_{AP}^{(2)}}\left[ {a,:} \right] \odot {\textbf{M}_{MP}}\left[ {m,:} \right] \right \|} }},
    \end{split}
    \label{}
    \end{equation}
    
    which can be interpreted as
    \begin{equation}
    \begin{split}
    &HeteAlloc(a,m) =\\
    &\frac{N_{\textup{papers of }a \textup{ which include  
     }m}}{\sqrt{N_{\textup{papers of }a}}\cdot \sqrt{N_{\textup{papers of a's co-authors which include 
     }m}}}. 
     \end{split}
    \end{equation}
    
The advantage of this selection strategy is that the similarity between an author and any $MeSH$ term will not be influenced by an irrelevant global change of the data set. The subset matrix is constant for all target $MeSH$ terms. However, this selection does not reflect on which specific \textcolor{black}{$MeSH$ term} an author has collaborated with another author, and simply includes the papers of all co-authors into the subset.
    
\subsection{Subset of co-authors' papers in \textcolor{black}{a} target \textcolor{black}{$MeSH$ term}.} The basic idea of this strategy is to add the target $MeSH$ term as another constraint for selecting the subset. The subset includes all papers published by the target author and by the authors who have co-authored with him or her in the target \textcolor{black}{$MeSH$ term}. Since this subset varies according to $MeSH$ terms, we use the reachable vector of $a$ and $m$ to replace $\textbf{RM}_{sub} [a,:]$
    
    \begin{equation}
    \begin{split}
        &{\rm{HeteAlloc(a}},{\rm{\;m }}\left| {{\rm{ a\;}} \in {\rm{A}},{\rm{\;m\;}} \in {\rm{M}}} \right.{\rm{ }}) =\\
        &\frac{{\textbf{M}_{AP}}\left[ {a,:} \right] \cdot ({\textbf{RV}_{sub}^{(a,m)}} \odot \textbf{M}_{MP}\left[ {m,:} \right])'}{{\sqrt  {\left \|  {\textbf{M}_{AP}}\left[ {a,:} \right] \right \|} \; \cdot \;\sqrt {\left \| {\textbf{RV}_{sub}^{(a,m)}}\odot {\textbf{M}_{MP}}\left[ {m,:} \right] \right \|} }}
        \end{split}
        \label{aa}
    \end{equation}
    
    \begin{equation}
        \textbf{RV}_{Sub}^{(a,m)}\left( {1,n} \right)  = \begin{cases}
    0 & \text{ if } {\textbf{V}_{Sub}^{(a,m)}\left( {1,n} \right) = 0} \\ 
    1 & \text{ otherwise }  
    \end{cases} 
    \end{equation}
    
    \noindent where
    \begin{equation}
            \textbf{V}_{sub}^{(a,m)} = {\rm{\;}}({\textbf{W}_{AP}}\left( {a,:} \right) \odot {\textbf{W}_{MP}}\left( {m,:} \right)) \cdot {\textbf{W}_{PA}} \cdot {\textbf{W}_{AP}}.
    \end{equation}

   Equation~\ref{aa} can be interpreted as
     \begin{equation}
    \textcolor{black}{HeteAlloc(a,m) =  \frac{N_{\textup{papers of }a \textup{which include  
     }m}}{\sqrt{N_{\textup{papers of }a}}\cdot \sqrt{N_{\textup{papers of $a$'s co-authors which include m}}}}}. 
    \end{equation}
    
The advantage of this selection strategy is that the similarity between an author and any $MeSH$ term will not be influenced by any irrelevant global changes of the data set. The similarity is $MeSH$-sensitive, and the subset vector can filter out co-authors who had no experience on the target \textcolor{black}{$MeSH$ term}. However, this selection will lead to a low score for those who have worked with very experienced authors.
    
\subsection{Subset of all papers published by the co-authors of the focal paper.}

For each paper $p$, the subset includes all papers published by the co-authors of $p$. And for each pair, author $a$ and $MeSH$ term $m$, the calculation is conducted for every paper $p$ of author $a$ which includes the $MeSH$ term $m$, and the average or the sum of all papers is used as the final score. The sum can be considered as a method for credit allocation and the average as a similarity measure. Here we shall use the sum as an example:
    
    \begin{equation}
        HeteAlloc(a,m) = \sum_{p\in P_{a}} HeteAlloc(a,p,m)
    \end{equation}
    
    \begin{equation}
             HeteAlloc(a,p,m) = \frac{{\textbf{M}_{AP}}\left[ {a,:} \right] \cdot ({\textbf{RV}_{sub}^{(a,p)}} \odot \textbf{M}_{MP}\left[ {m,:} \right])'}{{\sqrt  {\left \|  {\textbf{M}_{AP}}\left[ {a,:} \right] \right \|} \; \cdot \;\sqrt {\left \| {\textbf{RV}_{sub}^{(a,p)}}\odot {\textbf{M}_{MP}}\left[ {m,:} \right] \right \|} }}
             \label{bb}
    \end{equation}
    
    \begin{equation}
        \textbf{RV}_{Sub}^{(a,p)}\left( {1,n} \right)  = \begin{cases}
    0 & \text{ if } {\textbf{V}_{Sub}^{(a,p)}\left( {1,n} \right) = 0} \\ 
    1 & \text{ otherwise }  
    \end{cases} 
    \end{equation}
    
    \noindent where
    \begin{equation}
            \textbf{V}_{sub}^{(a,p)} = {\rm{\;}}{\textbf{W}_{AP}}\left( {a,:} \right) \odot   {\textbf{W}_{PA}} \cdot {\textbf{W}_{AP}}(p,:).
    \end{equation}
    
   Equation~\ref{bb} can be interpreted as:
     \begin{equation}
     \begin{split}
    &HeteAlloc(a,m) =\\
    &\sum_{\textup{all papers of a}} \frac{N_{\textup{papers of }a \textup{which include  
     }m}}{\sqrt{N_{\textup{papers of }a}}\cdot \sqrt{N_{\textup{papers of co-authors of paper p}}}}.
    \end{split}
    \end{equation}
    
This similarity avoids a significant decrease when the target author co-authors with a more experienced one in the target \textcolor{black}{$MeSH$ term}. The similarity retains the property of having a $MeSH$-sensitive subset. Notice that this method works better when applied to calculate the absolute value of expertise.

\section{Extensions of \textit{HeteAlloc}}
\label{sec:ext}
\subsection{Weighted version of \textit{HeteAlloc}\label{weighted}}

\textcolor{black}{The formalisation above is based on an unweighted network. Yet, one may want to capture the concentration of an author's effort on a specific topic ($MeSH$ term). For example, let us suppose that all papers of author $A_1$ only contain one $MeSH$ term $M1$ and all papers of another author $A_2$ contain two $MeSH$ terms, $M1$ and $M2$. In this case, one may argue that $A_1$ concentrates more than $A_2$ on $M1$ since $A_1$ has worked exclusively on this topic while $A_2$ on the additional topic $M2$.}
According to this idea, we propose a weighted version of \textit{HeteAlloc} which accounts for the weights of the links between papers and $MeSH$ terms. \textcolor{black}{The weight of a link between a paper and a $MeSH$ term is inversely proportional to the number of $MeSH$ terms associated with the paper.}
\textit{HeteAlloc} can be applied to a weighted network by using $U_{MP}$ instead of $M_{MP}$, where $U_{MP}$ is a normalised matrix of $M_{MP}$ along the column vector.

The weighted \textit{HeteAlloc} can capture authors' concentration on specific topics and identify the authors whose papers are more focused on smaller $MeSH$ sets. However, this characteristic is not necessarily an advantage, but simply a different strategy to deal with the number of $MeSH$ terms in a paper. There may exist different views about the similarity between an author and a given \textcolor{black}{$MeSH$ term}. For example, one may believe that an author is entirely devoted to a given \textcolor{black}{research topic}, if each of his or her papers contains the corresponding $MeSH$ term. In this case, the similarity between the author and the \textcolor{black}{$MeSH$ term} would be equal to one (i.e., the idea behind the unweighted version). However, others may believe that the similarity between the author and the \textcolor{black}{$MeSH$ term} should never be equal to $1$ unless an author’s work is exclusively about this $MeSH$ term (i.e., the idea behind the weighted version). The decision should be made after careful examination of the context, and should also be based on the assumptions made by \textcolor{black}{potential users of the method (e.g., researchers or funding agencies.)}. 

Here we shall provide our personal recommendation and blueprint. %
For smaller \textcolor{black}{$MeSH$ term} numbers, the weighted version will work better since it is not common for researchers to work in a completely different \textcolor{black}{$MeSH$ term} (say, Finance and Chemistry). However, when the division of topics is too fragmented and most papers have many $MeSH$ terms, then the performance of the weighted version may not work well, and the unweighted version would be recommended. 

\subsection{Iterative calculations over the years}
The original \textit{HeteSim} is designed for a ``static'' measurement of similarity. However, authors keep publishing papers over the years, and their expertise may change over time. When expertise is measured at year $T$, only the papers published before this year should be considered. To make our method \textit{HeteAlloc} applicable to dynamic calculation, we distinguish the links connecting Author and Paper between the experience/history links before year $T$ and the update links at year $T$. This can be done by using two adjacency matrices: $M_{update}$ and $M_{experience}$. Since it is difficult to identify the time ordering of publications published in the year $T$, we assume that papers of year $T$ were published at the same time. The formalisation of \textit{HeteAlloc} needs to be modified and the calculation, based on the modified measure, can be conducted iteratively over the years. 

We shall refer to the modified algorithm as \textit{DynamicHeteAlloc} (\textit{DHA}), and the corresponding formalisation is
\begin{equation}
    DHA(a,m) = \sum_{p_i\in M_{update}[a, :]\odot M_{MP} } DHA(a,p_i,m) 
\end{equation}
and
\begin{equation}
   \textcolor{black}{DHA(a,p_i,m) = \frac{(\textbf{M}_{experience}[a,:] + \textbf{I}_{nn}[p_i,:])\cdot (\textbf{V}_{subset}(p_i)\odot \textbf{M}_{MP}[m,:]) }{\sqrt{\left \|\textbf{M}_{experience}[a,:] + \textbf{I}_{nn}[p_i,:] \right \|\left \|\textbf{V}_{subset}(p_i)\odot \textbf{M}_{MP}[m,:] \right \|}}},
   \label{DHA}
\end{equation}

\noindent where

\begin{equation}
    \textbf{V}_{subset}(p_i) = \textbf{M}_{update}'[p_i,:]*\textbf{M}_{experience }+ \textbf{I}_{nn}'[p_i,:].
\end{equation}

For each paper, we add $\textbf{I}_{nn}[p_i,:]$ to $\textbf{M}_{experience}[a,:]$ in Equation~\ref{DHA} to include the current paper in the experience paper set so as to avoid the case where $\textbf{M}_{experience}$ is a zero matrix. 

According to the formalisation of \textit{DHA}, we have implemented Algorithm~\ref{Algo1}:

\begin{algorithm}[H]
    \caption{Algorithm for conducting dynamic \textit{HeteAlloc}}
    \begin{algorithmic}[1]
    
        \Require link lists for every year, $MeSH$ lists
        \Ensure expertise of every author
        \State initial $list_{pre}$ as blank list, load $MeSH$ list as $ \textbf{M}_{MP}$ ;
        \For{each $year \in [1946,2007]$}
            \State load $list_{year}$ as $list_{cur}$;
            \State Sparse matrix Creation; 
            \For{each $Author ID \in list_{cur}$}
                \If {$\textbf{M}_{update}[Author~ID,:]$ is Null vector}  
                    Next iteration;
                \EndIf
                \State find MeSH terms needed to update $MeSH_{update}$;
                \State create a null dictionary $dic_{cur}$;
                \If{Author ID exists in expertise dictionary $dic_{expts}$} use $dic_{expts}[Author~ID]$ to replace $dic_{cur}$
                \EndIf
                \For{each $MeSH~ID \in MeSH_{update}$}
                    Initialize \textit{HeteAlloc}\_value as zero;
                    \If{$MeSH$ ID in $dic_{cur}$ } use $dic_{cur}[$MeSH$~ID]$ to replace \textit{HeteAlloc}\_value
                    \EndIf
                    \State update \textit{HeteAlloc}\_value by adding result from Dynamic HeteAlloc(Author ID, $MeSH$ ID)
                    \State update $dic_{cur}[$MeSH$~ID]$ by \textit{HeteAlloc}\_value
                \State update $dic_{expts}[Author~ID]$ by $dic_{cur}$
            \EndFor
            
        \EndFor
    \EndFor
    \State Write out $dic_{expts}$.
    \end{algorithmic}
    \label{Algo1}
\end{algorithm}

\begin{algorithm}[H]
    \caption{Sparse Matrix Creation}
    \begin{algorithmic}[1]
        \Require $list_{pre}$, $list_{cur}$, $MeSH$ lists
        \Ensure  $\textbf{M}_{experience}$, $\textbf{M}_{update}$, update $list_{pre}$, dictionaries
        \State   merge $list_{pre}$ and $list_{cur}$ as $list_{all}$; 
        \State   create a dictionary from $list_{all}$ for mapping nodes with indexes;
        \State   use the dictionary to map $list_{pre}$ as $\textbf{M}_{experience}$, map $list_{cur}$ as $\textbf{M}_{update}$;
        \State   replace $list_{pre}$ by $list_{all}$, return dictionaries for mapping.
    \end{algorithmic}
\end{algorithm}

\textcolor{black}{
An example of this method using illustrative networks is provided in the Appendix. The results are given in the form of expertise matrices, where the value corresponding to row $i$ and column $j$ indicates the expertise of $Author_i$ on $MeSH_j$. In the example, we use the publication lists of $4$ authors from year $1$ to year $10$ and calculate the expertise matrices for each author at each year. We also show the result using the (\textit{BL}) method, which equally attributes every $MeSH$ term of a paper to all co-authors. In this case, the expertise of a focal author is therefore computed through the cumulative counts of $MeSH$ terms associated with all publications of the author. Thus, in the expertise matrix calculated using the (\textit{BL}) method for a year $t$, the value in row $i$ and column $j$ is equal to the number of papers published by $Author_i$ with $MeSH_j$ before year $t$.}

\section{Results}
\label{sec:results}
To compare the performance of different selections of subsets on HIN, we have calculated the similarity between all pairs extracted from the pair set $\left\{ a,m|a\in Author,m\in MeSH \right\}$ based on three small examples of networks using the (\textit{BL}) method mentioned above, the original \textit{HeteSim}, the \textit{HeteAlloc} with the subset of co-authors’ papers (\textit{HA1}), the \textit{HeteAlloc} with the subset of co-authors’ papers in \textcolor{black}{a} target \textcolor{black}{$MeSH$ term} (\textit{HA2}), the \textit{HeteAlloc} with the subset of all papers published by the co-authors of the focal paper (\textit{HA3}), and the corresponding weighted versions of \textit{HA1, HA2, HA3} (i.e., \textit{WHA1, WHA2, WHA3}).   

\begin{figure}[H]
\centering

\includegraphics[width=0.7\textwidth]{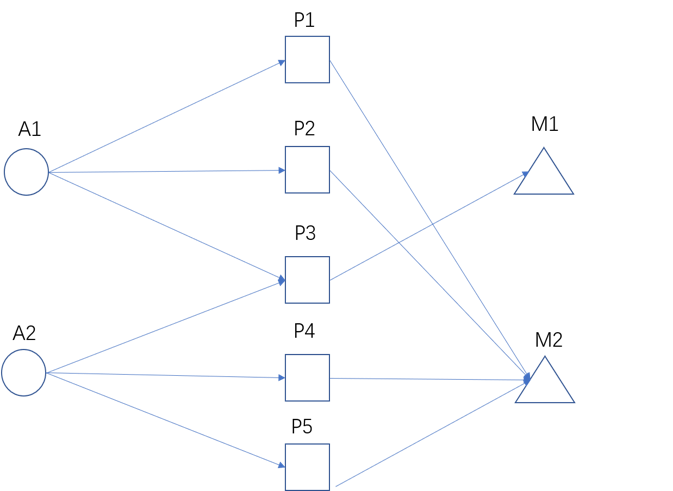}
\caption{Example network 1}
\label{fig:fig2}
\end{figure}

In the first example in Fig.~\ref{fig:fig2}, \textcolor{black}{\textit{BL}}, \textit{HA2} and \textit{HA3} perform well \textcolor{black}{(see Table \ref{tab:fig2}; the similarities characterised by better performance have been highlighted in bold)}. These methods can uncover the difference between $(A1, M1)$ and $(A1, M2)$. To be more specific, $A1$ published two papers with $M1$ and just one paper with $M2$, and the similarity between $A1$ and $M1$ should be higher than that between $A1$ and $M2$. Since each paper contains only one $MeSH$ term, the weighted versions in this example degenerate to the unweighted ones.

\begin{figure}[H]
\centering
\includegraphics[width=0.7\textwidth]{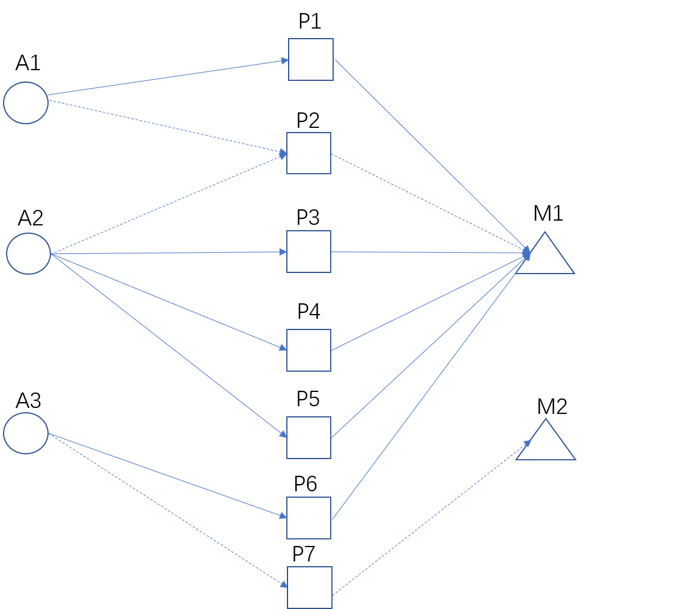}
\caption{Example network 2}
\label{fig:fig3}
\end{figure}

\begin{table}[htbp]
  \centering
 \caption{\textcolor{black}{Results based on example network 1}}
  \resizebox{\textwidth}{!}{
    \begin{tabular}{l|r|r|rrr|rrr}
          & \multicolumn{1}{c|}{Baseline} & \multicolumn{1}{c|}{Original} & \multicolumn{3}{c|}{Unweighted} & \multicolumn{3}{c}{Weighted} \\
    \hline
    \multicolumn{1}{c|}{Pair\textbackslash{}Method} & \multicolumn{1}{c|}{BL} &\multicolumn{1}{c|}{HeteSim} & \multicolumn{1}{c}{HA1} & \multicolumn{1}{c}{HA2} & \multicolumn{1}{c|}{HA3} & \multicolumn{1}{c}{WHA1} & \multicolumn{1}{c}{WHA2} & \multicolumn{1}{c}{WHA3} \\
    \hline
    (A1,M1) & 0.577 & 0.577 & 0.577 & 0.577 & 0.577 & 0.577 & 0.577 & 0.577 \\
    (A1,M2) & \textbf{0.816} & 0.577 & 0.577 & \textbf{0.816} & \textbf{0.816} & 0.577 & \textbf{0.816} & \textbf{0.816} \\
    (A2,M1) & 0.577 & 0.577 & 0.577 & 0.577 & 0.577 & 0.577 & 0.577 & 0.577 \\
    (A2,M2) & \textbf{0.816} & 0.577 & 0.577 & \textbf{0.816} & \textbf{0.816} & 0.577 & \textbf{0.816} & \textbf{0.816} \\
    \end{tabular}}%
  \label{tab:fig2}%
\end{table}%
In the second example network in Fig.~\ref{fig:fig3}, \textit{HA3} performs well. It shows that author $A1$ is more experienced than $A3$ in $M1$. To be more specific, $A1$ published a paper with $M1$ alone and another with \textcolor{black}{a} very experienced author, $A2$. $A3$ published a paper with $M1$ alone and another paper with $M2$ alone. The similarity between $A1$ and $M1$ should be greater than that between $A3$ and $M1$. Compared to other methods, only \textit{HA3} gives a higher similarity for $(A1, M1)$, and a higher score for the expert $A2$ with $M1$. Since each paper contains only one $MeSH$ term, the weighted versions in this example degenerate to the unweighted ones.

\begin{table}[htbp]
  \centering
  \caption{\textcolor{black}{Results based on example network 2}}
  \resizebox{\textwidth}{!}{
   \begin{tabular}{l|r|r|rrr|rrr}
          & \multicolumn{1}{c|}{Baseline} & \multicolumn{1}{c|}{Original} & \multicolumn{3}{c|}{Unweighted} & \multicolumn{3}{c}{Weighted} \\
    \hline
    \multicolumn{1}{c|}{Pair\textbackslash{}Method} & \multicolumn{1}{c|}{BL} &\multicolumn{1}{c|}{HeteSim} & \multicolumn{1}{c}{HA1} & \multicolumn{1}{c}{HA2} & \multicolumn{1}{c|}{HA3} & \multicolumn{1}{c}{WHA1} & \multicolumn{1}{c}{WHA2} & \multicolumn{1}{c}{WHA3} \\
    \hline
    (A1,M1) & 1 & 0.577 & 0.632 & 0.632 & \textbf{0.816} & 0.632 & 0.632 & \textbf{0.816} \\
    (A1,M2) & 0 & 0     & 0     & 0     & 0     & 0     & 0     & 0 \\
    (A2,M1) & 1 & 0.816 & 0.894 & 0.894 & \textbf{0.973} & 0.894 & 0.894 & \textbf{0.973} \\
    (A2,M2) & 0 & 0     & 0     & 0     & 0     & 0     & 0     & 0 \\
    (A3,M1) & 0.707 & 0.288 & 0.707 & 0.707 & \textbf{0.707} & 0.707 & 0.707 & \textbf{0.707} \\
    (A3,M2) & 0.707 & 0.707 & 0.707 & 0.707 & 0.707 & 0.707 & 0.707 & 0.707 \\
    \end{tabular}}%
  \label{tab:fig3}%
\end{table}%

\begin{figure}[H]
\centering
\includegraphics[width=0.7\textwidth]{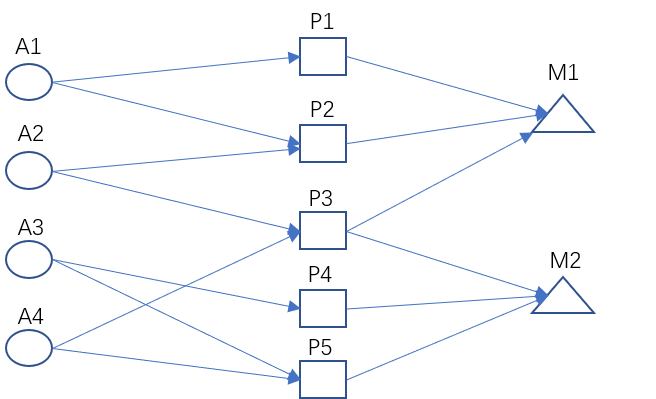}
\caption{Example network 3}
\label{fig:fig1}
\end{figure}

For the third example shown in Fig.~\ref{fig:fig1}, the weighted methods differentiate between $Sim(A1,M1)$ and $Sim(A2,M1)$, while the unweighted methods are unable to distinguish between them. To be more specific, both $A1$ and $A2$ published two papers with $M1$, and the only difference between $A1$ and $A2$ in $M1$ is that paper $P3$ published by $A2$ contains $M2$ as well. As mentioned in Section~\ref{weighted}, the weighted version can capture the concentration of research efforts in some \textcolor{black}{$MeSH$ terms}, and is biased in favour of the authors whose papers are more concentrated on a smaller $MeSH$ set.

\textcolor{black}{
\begin{table}[htbp]
  \centering
  \caption{\textcolor{black}{Results based on example network 3}}
  \resizebox{\textwidth}{!}{
    \begin{tabular}{l|r|r|rrr|rrr}
          & \multicolumn{1}{c|}{Baseline} & \multicolumn{1}{c|}{Original} & \multicolumn{3}{c|}{Unweighted} & \multicolumn{3}{c}{Weighted} \\
    \hline
    \multicolumn{1}{c|}{Pair\textbackslash{}Method} & \multicolumn{1}{c|}{BL} &\multicolumn{1}{c|}{HeteSim} & \multicolumn{1}{c}{HA1} & \multicolumn{1}{c}{HA2} & \multicolumn{1}{c|}{HA3} & \multicolumn{1}{c}{WHA1} & \multicolumn{1}{c}{WHA2} & \multicolumn{1}{c}{WHA3} \\
    \hline
    (A1,M1) & 1 & 0.943 & 0.816 & 0.816 & 0.908 & 0.943 & 0.943 & \textbf{0.971} \\
    (A1,M2) & 0 & 0     & 0     & 0     & 0     & 0     & 0     & 0 \\
    (A2,M1) & 0.816 & 0.707 & 0.816 & 0.816 & 0.908 & 0.707 & 0.707 & \textbf{0.828} \\
    (A2,M2) & 0.577 & 0.236 & 0.5   & 0.5   & 0.5   & 0.316 & 0.316 & 0.316 \\
    (A3,M1) & 0 & 0     & 0     & 0     & 0     & 0     & 0     & 0 \\
    (A3,M2) & 1 & 0.943 & 0.816 & 0.816 & 0.908 & 0.943 & 0.943 & \textbf{0.971} \\
    (A4,M1) & 0.577 & 0.236 & 0.5   & 0.5   & 0.5   & 0.316 & 0.316 & 0.316 \\
    (A4,M2) & 0.816 & 0.707 & 0.816 & 0.816 & 0.908 & 0.707 & 0.707 & \textbf{0.828} \\
    \end{tabular}}%
  \label{tab:fig1}%
\end{table}%
}

From the three examples above, the third subset selection strategy (i.e., subset of all papers published by the co-authors of the focal paper) outperforms the other two strategies. Moreover, by taking the sum of all scores (i.e., similarity measures) obtained from all publications of the focal author, this method enables us to evaluate the global expertise of an author based on his of her entire scientific production. 

In what follows, we will use the third selection strategy and perform a comparison between our method (\textit{DHA}) and the (\textit{BL}) method applied to the MEDLINE data set. \textcolor{black}{As in our data set most publications are associated with multiple $MeSH$ terms, we chose to use the unweighted version of our method.}

The output of both methods are vectors associated with authors representing their expertise in terms of each topic (i.e., $MeSH$ term). To compare the two methods, for each author we consider the following measures: (1) the ratio between maximum and minimum values of the author's expertise; (2) the author's maximum normalised expertise (i.e., obtained by dividing all values in a vector by its norm); and (3) the normalised maximum expertise of authors that have published more than $10$ papers at the time of the assessment of expertise (i.e., criterion 2 applied only to the subset of productive authors). Moreover, for every year, we calculate the mean and standard deviation of the values produced by the above assessment measures, and compare them between methods.

\begin{table}[htbp]
\centering
\begin{threeparttable}{
  \centering
  \caption{Comparison between \textit{DHA} and \textit{BL} based on the first 10 years of the MEDLINE data set}
 \begin{tabular}{l|c|cc|cc|cc}

     & \multicolumn{1}{c|}{Measure} & \multicolumn{2}{c|}{(1)} & \multicolumn{2}{c|}{(2)} & \multicolumn{2}{c}{(3)}\\
\hline
   year  & method &        DHA &        BL &            DHA &        BL &                 DHA &        BL \\
\hline
1948 & mean &    2.05 &  1.45 &           0.60 &  0.58 &                0.57 &  0.52 \\
     & std &    3.54 &  1.13 &           0.17 &  0.17 &                0.14 &  0.12 \\
1949 & mean &    2.72 &  1.66 &           0.60 &  0.58 &                0.59 &  0.54 \\
     & std &    6.24 &  1.63 &           0.16 &  0.16 &                0.14 &  0.12 \\
1950 & mean &    3.48 &  1.84 &           0.60 &  0.57 &                0.60 &  0.55 \\
     & std &    9.59 &  2.09 &           0.16 &  0.15 &                0.14 &  0.12 \\
1951 & mean &    4.37 &  2.06 &           0.59 &  0.56 &                0.61 &  0.56 \\
     & std &   13.85 &  2.65 &           0.15 &  0.14 &                0.14 &  0.12 \\
1952 & mean &    5.22 &  2.24 &           0.59 &  0.56 &                0.61 &  0.56 \\
     & std &   18.36 &  3.15 &           0.15 &  0.14 &                0.14 &  0.12 \\
1953 & mean &    6.05 &  2.39 &           0.59 &  0.55 &                0.61 &  0.56 \\
     & std &   23.02 &  3.60 &           0.15 &  0.14 &                0.14 &  0.11 \\
1954 & mean &    6.85 &  2.53 &           0.59 &  0.55 &                0.61 &  0.56 \\
     & std &   28.05 &  4.01 &           0.15 &  0.13 &                0.14 &  0.11 \\
1955 & mean &    7.65 &  2.66 &           0.59 &  0.54 &                0.61 &  0.55 \\
     & std &   33.04 &  4.41 &           0.15 &  0.13 &                0.14 &  0.11 \\
1956 & mean &    8.41 &  2.78 &           0.59 &  0.54 &                0.61 &  0.55 \\
     & std &   38.16 &  4.79 &           0.15 &  0.13 &                0.14 &  0.11 \\
1957 & mean &    9.14 &  2.88 &           0.59 &  0.54 &                0.61 &  0.55 \\
     & std &   43.32 &  5.13 &           0.15 &  0.13 &                0.14 &  0.11 \\
\bottomrule
\end{tabular}%
\label{tab:compare}
    \begin{tablenotes}
      \small
      \item (1) the ratio between maximum and minimum values of the author's expertise; (2) the author's maximum normalised expertise (i.e., obtained by dividing all values in a vector by its norm); and (3) the normalised maximum expertise of authors that have published more than $10$ papers at the time of the assessment of expertise
    \end{tablenotes}
    }
    
\end{threeparttable}
\end{table}%

\begin{figure}[htp]
    \centering
    \includegraphics[width=0.5\textwidth]{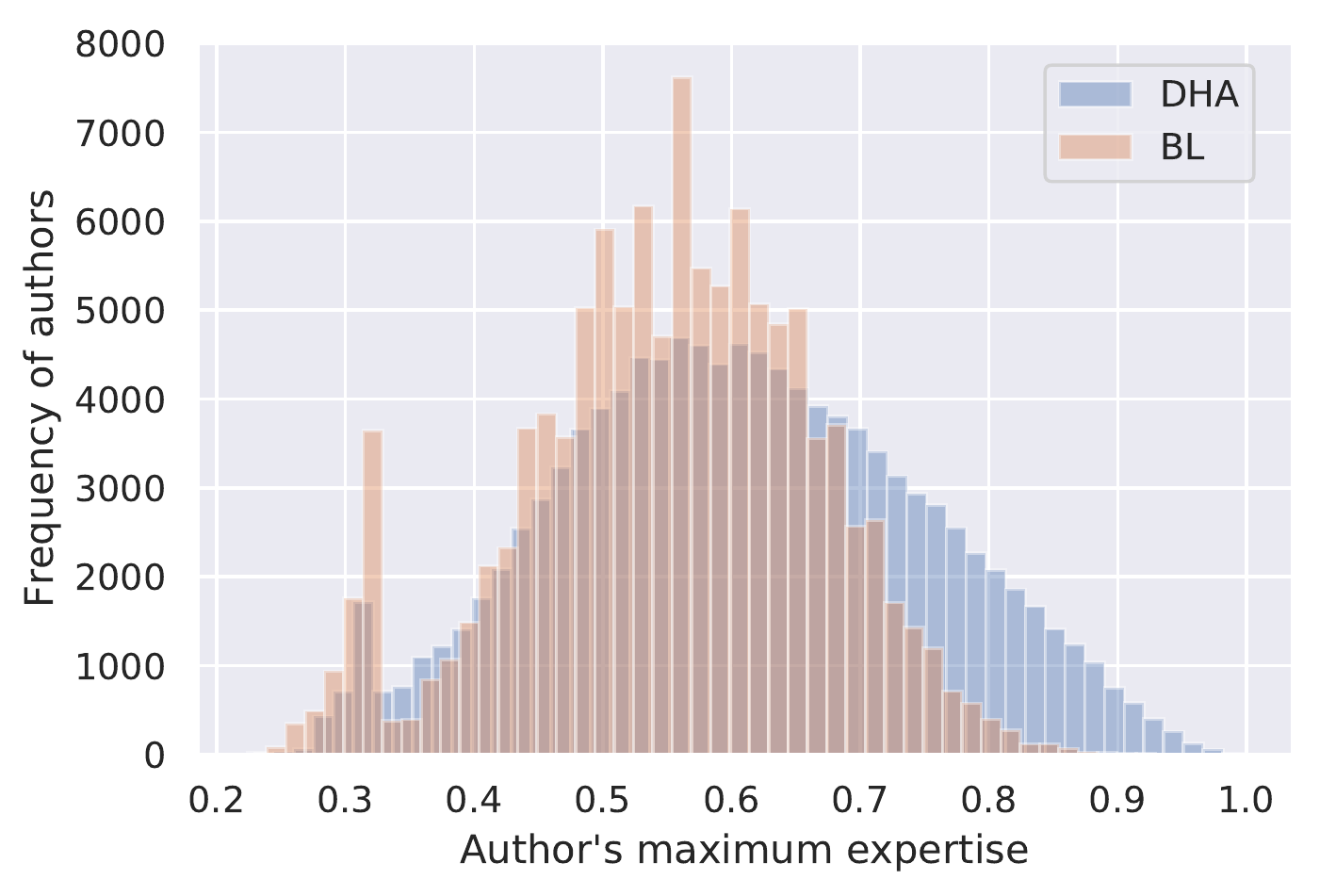}
    \caption{Comparison between \textit{DHA} and \textit{BL} using the normalised maximum expertise of productive authors}
    \label{fig:compare}
\end{figure}

The results reported in Table~\ref{tab:compare} show that the mean and standard deviation of the ratio between maximum and minimum values of author's expertise obtained with the \textit{DHA} method are higher than the mean and standard deviation obtained with the \textit{BL} method, which suggests that \textit{DHA} can better distinguish authors according to their expertise areas, whereas $BL$ considers all authors involved in works relevant to multiple topics as interdisciplinary authors (i.e., with the same expertise on all \textcolor{black}{$MeSH$ terms}, thus producing smaller ratios of maximum to minimum values of expertise). The results based on normalised maximum expertise of \textit{DHA} are similar to those of \textit{BL} when all authors are considered, but they differ when the methods are applied only to \textcolor{black}{a} restricted subset of productive authors, which suggests that our method has the potential to identify authors' main areas of expertise precisely when they are most likely to work in multiple areas. 

Figure~\ref{fig:compare} shows the frequency of productive authors with normalised maximum expertise ranging from $0$ to $1$. The (\textit{BL}) method shows no authors with maximum expertise higher than $0.9$, which suggests that there is no researcher dedicated to one single area and the maximum expertise of most authors lies in the middle. However, the results obtained with our method clearly highlight its ability to identify specialised authors that preferentially focus on one area (i.e., with high maximum expertise) and at the same time \textcolor{black}{interdisciplinary authors whose work spans different areas} (i.e., those with low maximum expertise).

\section{Conclusions\label{sec:conclusion}}

In this work, we have proposed a new method based on path similarity and a number of subset selection strategies to identify authors' expertise. %
Our method differs from previous works as it assigns expertise to a focal author by accounting for co-authors' contributions to the works they were involved with. We have shown that our method can be applied to the HIN constructed from the MEDLINE corpus. However, the applicability of our method is not limited to just one data set. Indeed if we replace $MeSH$ terms by the topic tags in MAG, our method can be directly applied to MAG. In this case, it can retrieve authors' expertise based on topics as classified in MAG, and it can be suitably adjusted to reflect the depth and granularity required by users. In more general cases, users can generate their own topics from documents using topic modelling or other methods. By linking the generated topics and the corresponding documents, users can produce similar networks as those shown in Fig.~\ref{fig:HIN_demo} and they can then apply our method by selecting an appropriate subset. Our work can also be used to integrate standard approaches, for example in conjunction with topic modelling for documents or by using topic classification systems. 

\textcolor{black}{The lack of a ground truth does not enable a definitive validation of our method. While this represents a limitation of our work, it also opens up new avenues for future work. For example, to mitigate this limitation, we could check the Contributor Roles Taxonomy (CRediT) author statement available from several journals\footnote{https://www.elsevier.com/authors/journal-authors/policies-and-ethics/credit-author-statement} to identify which author was involved in which part of the research. However, CRediT statements are self-declared and not verifiable, which again highlights the need for methods such as the one we proposed in this article. Moreover, the CRediT author statements are not detailed enough to unambiguously indicate which specific expertise (e.g., $MeSH$ term) should be associated with which author. Another possibility is to handpick some very interdisciplinary papers (i.e., with many $MeSH$ terms). By reading the CV of the authors or searching for relevant information about them, we might be able to infer the $MeSH$ terms associated with each author, and then compare our prior knowledge with the results obtained using our method. This test represents a ``sanity check'', and an example is given in the Appendix.
}

\textcolor{black}{Our method has a number of important applications for research and practice. 
Understanding the composition of a team and being able to associate each co-author of a paper to one or several fields of expertise can spur new studies of the interdisciplinarity of research teams. For example, our method will enable us to distinguish between interdisciplinary papers co-authored by researchers with overlapping expertise, and equally interdisciplinary papers in which the co-authors have non-overlapping research profiles. This, in turn, could shed further light on the impact of team diversity on scientific success and knowledge creation. Moreover, being able to identify expertise facilitates a comparative assessment of two equally interdisciplinary studies, one pursued by an individual and the other by a group or researchers. In particular, our method enables us to distinguish between research solely pursued by one individual scholar with a highly interdisciplinary background and research pursued by an interdisciplinary group comprising of several highly specialised scholars. This variation in type and sources of interdisciplinarity is likely to be a critical nuance with non-trivial implications for innovation, research performance, and the long-term impact of publications. }

\textcolor{black}{Our method has also practical implications for funding agencies, research institutions and scientists. First, it can assist funding agencies in the identification of appropriate reviewers with the right competence to evaluate research proposals. In turn, it may also assist reviewers in uncovering possible gaps between a proposed research and the combined expertise of the pool of applicants. Second, our method can also help research institutions to develop effective recruitment policies targeted at strengthening specific research fields or at developing new and fast-developing areas that require a prompt investment of resources. Finally, the identification of special expertise can help scientists in identifying potential collaborators and shaping successful research groups.}

\appendix
\color{black}
\section{\textcolor{black}{Appendix}}\label{sec:Appendix}

\subsection{Example of \textit{DHA} using illustrative networks}\label{sec:A1}

\begin{figure}[H]
    \centering
    \includegraphics[width=0.8\textwidth]{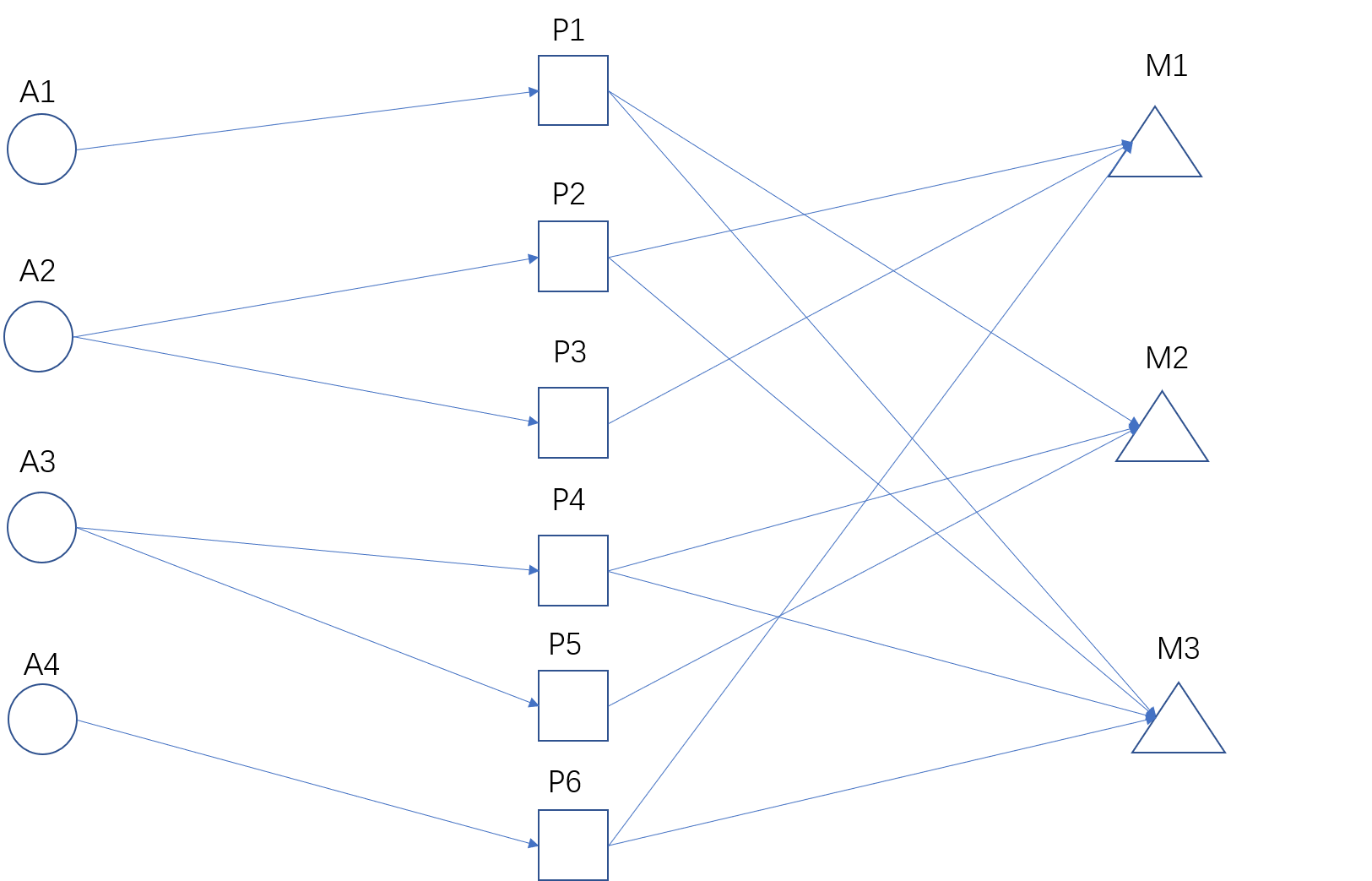}
    \caption{\color{black}Illustrative networks from year $1$ to year $5$}
    \label{fig:toy1}
\end{figure}

\begin{figure}[H]
    \centering
    \includegraphics[width=0.8\textwidth]{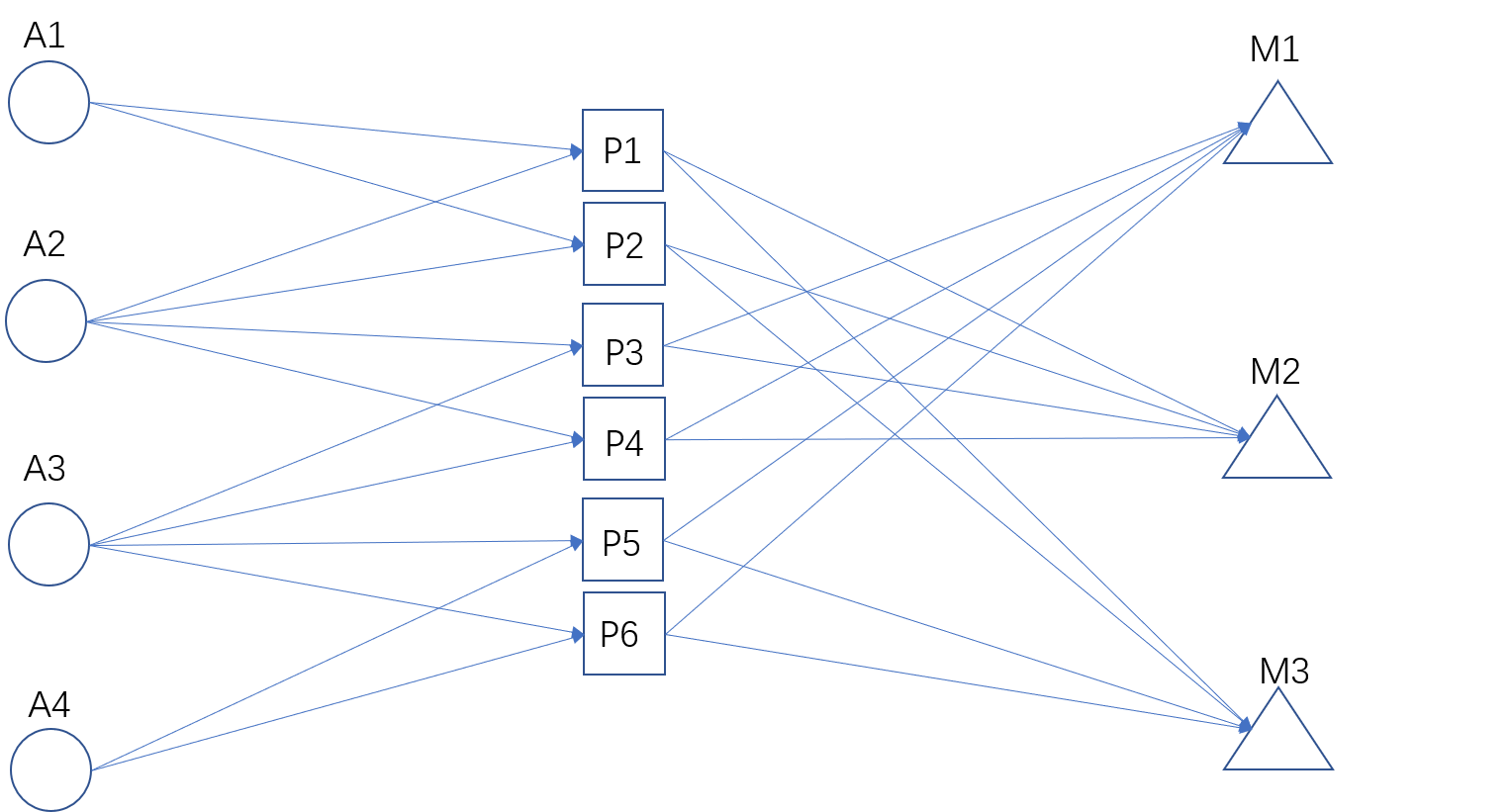}
    \caption{\color{black}Illustrative networks from year $6$ to year $10$}
    \label{fig:toy2}
\end{figure}

Here we show how our method works out in full using illustrative networks, and we then compare the results with those obtained using the $BL$ method. Figure~\ref{fig:toy1} shows the illustrative networks from year $1$ to year $5$ (identical networks for five years). Figure~\ref{fig:toy2} shows the illustrative networks from year $6$ to year $10$ (identical networks for five years). Before year $5$, the four authors worked separately. $A1$ worked on $M2$ and $M3$ equally. $A2$ mainly worked on $M1$ and had some works related to $M3$. $A3$ mainly worked on $M2$ and had some works related to $M3$. $A4$ worked on $M1$ and $M3$ equally. From year $6$, they started to collaborate. Specifically, $A1$ and $A2$ collaborated on papers related to $M2$ and $M3$, $A2$ and $A3$ collaborated on $M1$ and $M2$, $A3$ and $A4$ collaborated on $M1$ and $M3$. The publication lists can be found in Tables \ref{tab:pub_list} and \ref{tab:pub_list1}. 

Based on their experience, it is not likely for $A2$ to have many contributions on $M2$ in $P1$ from year $6$ to year $10$ since he or she did not have any previous experience on that $MeSH$ category. Similarly, it is not likely for $A3$ to have many contributions on $M1$ in $P2$ from year $6$ to year $10$. But they may acquire some experience from those collaborations. Thus, a good method should be able to allocate the credit of those collaborative works to those collaborators with corresponding experience.

Equations \ref{equation:year1}--\ref{equation:year6} listed the expertise matrices given by \textit{BL} and \textit{DHA}, respectively. The results are similar between year $1$ and year $5$ and begin to differentiate from year $6$. 

At the end of year $5$, both methods suggest that all four authors had similar expertise on $M3$, whereas $A2$ and $A3$ were experts on $M1$ and $M2$, respectively. $BL$ simply counts for the number of papers each author published on every $MeSH$ term, and adds them together. Following this idea, $A2$ gained the same amount of credit as $A1$ on $M2$ from $P1$ and as $A3$ from $P2$ from year $6$ to year $10$ although $A2$ never worked on $M2$ before year $6$. As a result, at the end of year $10$, $A2$ was recognised as an expert on $M2$, with the same expertise as $A3$. 

However, under most circumstances, the contribution each scholar makes to the joint work is likely to relate to the specific topics or fields in which his or her expertise lies. %
Specifically, it is more reasonable to think that during the collaboration of $P2$ from year $6$ to year $10$, $A2$ contributed on $M1$ and $A3$ contributed on $M2$ based on their expertise. Therefore, $A2$ should gain the credit of $M1$ and $A3$ should gain the credit of $M2$. And the results obtained using \textit{DHA} gave the expected result: i.e., $A2$ is an expert on $M1$ and $A3$ is an expert on $M2$.

\begin{table}[H]
  \centering
  \caption{\color{black}Publication list in the illustrative networks from year $1$ to year $5$}
  {\color{black}
    \begin{tabular}{llll}
    \hline
    Author & Paper & MeSH  & Year \\
    \hline
    A1    & P1    & M2,M3 & 1,2,3,4,5 \\
    A2    & P2    & M1,M3 & 1,2,3,4,5 \\
    A2    & P3    & M1    & 1,2,3,4,5 \\
    A3    & P4    & M2,M3 & 1,2,3,4,5 \\
    A3    & P5    & M2    & 1,2,3,4,5 \\
    A4    & P6    & M1,M3 & 1,2,3,4,5 \\
    \hline

    \end{tabular}%
}
  \label{tab:pub_list}%
\end{table}%

\begin{table}[H]
  \centering
  \caption{\color{black}Publication list in the illustrative networks from year $6$ to year $10$}
      {\color{black}     \begin{tabular}{llll}
    \hline
    Author & Paper & MeSH  & Year \\
    \hline
    A1, A2 & P1    & M2,M3 & 6,7,8,9,10 \\
    A1, A2 & P1    & M2,M3 & 6,7,8,9,10 \\
    A2, A3 & P2    & M1,M2 & 6,7,8,9,10 \\
    A2, A3 & P2    & M1,M2 & 6,7,8,9,10 \\
    A3, A4 & P3    & M1,M3 & 6,7,8,9,10 \\
    A3, A4 & P3    & M1,M3 & 6,7,8,9,10 \\
    \hline

    \end{tabular}}%
  \label{tab:pub_list1}%
\end{table}%

\begin{equation}\label{equation:year1}
\textbf{M}_{t_1}^{BL}={
\left[ \begin{array}{ccc}
0 & 1 & 1\\
2 & 0 & 1\\
0 & 2 & 1\\
1 & 0 & 1
\end{array} 
\right ]},
\textbf{M}_{t_1}^{DHA}={
\left[ \begin{array}{ccc}
0 & 1 & 1\\
2 & 0 & 1\\
0 & 2 & 1\\
1 & 0 & 1
\end{array}
\right ]}
\end{equation}

\begin{equation}\label{equation:year2}
\textbf{M}_{t_2}^{BL}={
\left[ \begin{array}{ccc}
0 & 2 & 2\\
4 & 0 & 2\\
0 & 4 & 2\\
2 & 0 & 2
\end{array} 
\right ]},
\textbf{M}_{t_2}^{DHA}={
\left[ \begin{array}{ccc}
0 & 1.95 & 1.95\\
3.95 & 0 & 1.84\\
0 & 3.95 & 1.84\\
3.95 & 0 & 1.84
\end{array}
\right ]}
\end{equation}

\begin{equation}\label{equation:year3}
\textbf{M}_{t_3}^{BL}={
\left[ \begin{array}{ccc}
0 & 3 & 3\\
6 & 0 & 3\\
0 & 6 & 3\\
3 & 0 & 3
\end{array} 
\right ]},
\textbf{M}_{t_3}^{DHA}={
\left[ \begin{array}{ccc}
0 & 2.86 & 2.86\\
5.88 & 0 & 2.61\\
0 & 5.88 & 2.61\\
2.86 & 0 & 2.86
\end{array}
\right ]}
\end{equation}

\begin{equation}\label{equation:year4}
\textbf{M}_{t_4}^{BL}={
\left[ \begin{array}{ccc}
0 & 4 & 4\\
8 & 0 & 4\\
0 & 8 & 4\\
4 & 0 & 4
\end{array} 
\right ]},
\textbf{M}_{t_4}^{DHA}={
\left[ \begin{array}{ccc}
0 & 3.76 & 3.76\\
7.82 & 0 & 3.33\\
0 & 7.82 & 3.33\\
3.76 & 0 & 3.76
\end{array}
\right ]}
\end{equation}

\begin{equation}\label{equation:year5}
\textbf{M}_{t_5}^{BL}={
\left[ \begin{array}{ccc}
0 & 5 & 5\\
10 & 0 & 5\\
0 & 10 & 5\\
5 & 0 & 5
\end{array} 
\right ]},
\textbf{M}_{t_5}^{DHA}={
\left[ \begin{array}{ccc}
0 & 4.64 & 4.64\\
9.75 & 0 & 4.02\\
0 & 9.75 & 4.02\\
4.64 & 0 & 4.64
\end{array}
\right ]}
\end{equation}

\begin{equation}\label{equation:year6}
\textbf{M}_{t_6}^{BL}={
\left[ \begin{array}{ccc}
0 & 7 & 7\\
12 & 4 & 7\\
4 & 12 & 7\\
7 & 0 & 7
\end{array} 
\right ]},
\textbf{M}_{t_6}^{DHA}={
\left[ \begin{array}{ccc}
0 & 5.94 & 5.79\\
11.15 & 0.34  & 4.86\\
0.34  & 11.15 & 4.86\\
5.94 & 0 & 5.79
\end{array}
\right ]}
\end{equation}

\begin{equation}\label{equation:year7}
\textbf{M}_{t_7}^{BL}={
\left[ \begin{array}{ccc}
0 & 9 & 9\\
14 & 8 & 9\\
8 & 14 & 9\\
9 & 0 & 9
\end{array} 
\right ]},
\textbf{M}_{t_7}^{DHA}={
\left[ \begin{array}{ccc}
0 & 7.22 & 6.93\\
12.54 & 0.72  & 5.69\\
0.72 & 12.54 & 5.69\\
7.22 & 0 & 6.93
\end{array}
\right ]}
\end{equation}

\begin{equation}\label{equation:year8}
\textbf{M}_{t_8}^{BL}={
\left[ \begin{array}{ccc}
0 & 11 & 11\\
16 & 12 & 11\\
12 & 16 & 11\\
11 & 0 & 11
\end{array} 
\right ]},
\textbf{M}_{t_8}^{DHA}={
\left[ \begin{array}{ccc}
0 & 8.50 & 8.05\\
13.92 & 1.14  & 6.52\\
1.14 & 13.92 & 6.52\\
8.50 & 0 & 8.05
\end{array}
\right ]}
\end{equation}

\begin{equation}\label{equation:year9}
\textbf{M}_{t_9}^{BL}={
\left[ \begin{array}{ccc}
0 & 13 & 13\\
18 & 16 & 13\\
16 & 18 & 13\\
13 & 0 & 13
\end{array} 
\right ]},
\textbf{M}_{t_9}^{DHA}={
\left[ \begin{array}{ccc}
0 & 9.76 & 9.15\\
15.3 & 1.61 & 7.33\\
1.61 & 15.3 & 7.33\\
9.76 & 0 & 9.15
\end{array}
\right ]}
\end{equation}

\begin{equation}\label{equation:year10}
\textbf{M}_{t_{10}}^{BL}={
\left[ \begin{array}{ccc}
0 & 15 & 15\\
20 & 20 & 15\\
20 & 20 & 15\\
15 & 0 & 15
\end{array} 
\right ]},
\textbf{M}_{t_{10}}^{DHA}={
\left[ \begin{array}{ccc}
0 & 11.02 & 10.25\\
\textbf{16.66} & \textbf{2.12}  & 8.14\\
\textbf{2.12} & \textbf{16.66} & 8.14\\
10.25 & 0 & 11.02
\end{array}
\right ]}
\end{equation}

\subsection{An example based on real data}\label{sec:A2}
Here we provide an example based on a focal paper and show the results obtained using our method. 
The title of this focal paper is: "Calcium Levels and Calciuria in Decalcification in Acromegaly"
\footnote{https://pubmed.ncbi.nlm.nih.gov/13327374/}. 
It was published in 1956, co-authored by five authors: S. De S\`{e}ze, A. Lichtwitz, D. Hioco, M. Delaville, H. Gille. 
Table \ref{tab:$MeSH$_ex} shows the $MeSH$ terms associated with this paper, the relevant $MeSH$ Tree ID and the corresponding category names. 
Table \ref{exp_before} shows the expertise of the five co-authors on the $MeSH$ terms associated with the focal paper before the year 1956. 
The first author, Stanislas de S\`{e}ze, was a pioneering scholar of French rheumatology. He was already an expert in two categories: Musculoskeletal Disease, Nervous System Diseases and Humans (included in the category Eukaryota). This was indicated by the high values in his expertise vector: $90$ for B01, $42$ for C05 and $12$ for C10. 
The second author, Alfred Lichtwitz, mainly worked on D06, B01 and C19. The third author, Denis Hioco, mainly worked on D01, D06, A12. The fourth author, M.Delaville, mainly worked on B01, D06, D01. The last author, Halvor Gille, was a new author, and this paper was his first publication. 

Although there were some overlaps among those co-authors' profiles, each of those co-authors (except the new author) had some major background knowledge in selected research areas. The desired method should be able to add appropriate value to the co-authors' expertise vectors and update the expertise vectors so that they can better represent the evolution of the co-authors' expertise.    

The results are given in Table \ref{tab:exp_acq}. Upon publication of this paper, Stanislas de S\`{e}ze obtains $0.762$ on B01, $0.371$ on C05 and $0.106$ on C10, since he was the most experienced author in these three categories. Similarly, D. Hioco obtains $0.315$ on D01 and $0.265$ on A12; A. Lichtwitz obtains $0.193$ on D01 and $0.211$ on C19. However, M. Delaville does not achieve a high score as he was not the most experienced author in any of these categories. As for the new author, he gains some experience in nearly every category, especially those in which no one had much experience. In this example, he obtained $0.535$ on D23, $0.424$ on G02 and $0.366$ on G03. In general, our method clearly returns a reasonable result which meets our expectation. 

\begin{table}[htb]
\centering
\caption{\color{black}$MeSH$ terms associated with the focal paper, relevant $MeSH$ Tree ID and corresponding category names}
\label{tab:$MeSH$_ex}
 \resizebox{\textwidth}{!}{
  {\color{black}     \begin{tabular}{l|l|l}

\hline
$MeSH$ term  &  Relevant $MeSH$ Tree ID  &  Categories   \\
\hline

Acromegaly & [C05,C10,C19] & [Musculoskeletal Diseases; Nervous System Diseases; Endocrine System Diseases] \\

Calcium & [D01,D23] & [Inorganic Chemicals; Biological Factors]\\

Hormones & [D06,D27] & [Hormones, Hormone Substitutes, and Hormone Antagonists; Chemical Actions and Uses]
\\

Humans & [B01] & [Eukaryota]
\\
Osteoporosis & [C05,C18] & [Musculoskeletal Diseases; Nutritional and Metabolic Diseases]
\\
Phosphorus & [D01] & [Inorganic Chemicals]
\\
Urine & [A12] & [Fluids and Secretions]
\\
Water-Electrolyte Balance & [G02,G03,G07] & [Chemical Phenomena; Metabolism; Physiological Phenomena ]
\\
\hline
\end{tabular}}
}
\end{table}

\begin{table}[htb]
\centering
\caption{\color{black}Expertise of co-authors on the $MeSH$ terms associated with the focal paper before year 1956}
\label{exp_before}
 \resizebox{\textwidth}{!}{
   {\color{black}     \begin{tabular}{lrrrrrrrrrrrrr}
\toprule
{} &    D06 &    D27 &     B01 &    D01 &    D23 &    A12 &     C05 &     C10 &    C19 &    G02 &    G03 &    G07 &    C18 \\
\midrule
M. Delaville &  4.860 &  2.477 &  10.200 &  3.235 &  0.188 &  0.915 &   1.472 &   0.211 &  2.758 &  0.456 &  0.089 &  1.010 &  0.971 \\
H. Gille &  0.000 &  0.000 &   0.000 &  0.000 &  0.000 &  0.000 &   0.000 &   0.000 &  0.000 &  0.000 &  0.000 &  0.000 &  0.000 \\
A. Lichtwitz  &  7.139 &  3.963 &  22.821 &  3.141 &  1.295 &  0.987 &   4.754 &   1.064 &  6.219 &  2.074 &  2.406 &  2.576 &  2.821 \\
D. Hioco &  3.283 &  2.543 &   1.172 &  3.887 &  0.863 &  2.338 &   0.444 &   0.289 &  0.973 &  0.000 &  0.816 &  0.131 &  2.014 \\
De S\`{e}ze &  3.514 &  0.682 &  90.230 &  0.417 &  0.196 &  0.157 &  42.682 &  12.108 &  0.390 &  0.213 &  0.000 &  0.133 &  0.697 \\
\bottomrule
\end{tabular}}
}
\end{table}

\begin{table}[htb]
\centering
\caption{\color{black}Expertise acquired from the focal paper}
\label{tab:exp_acq}
 \resizebox{\textwidth}{!}{
  {\color{black}     \begin{tabular}{lrrrrrrrrrrrrr}
\toprule
{} &    D06 &    D27 &    B01 &    D01 &    D23 &    A12 &    C05 &    C10 &    C19 &    G02 &    G03 &    G07 &    C18 \\
\midrule
M. Delaville &  0.185 &  0.097 &  0.084 &  0.141 &  0.041 &  0.048 &  0.005 &  0.006 &  0.092 &  0.038 &  0.027 &  0.051 &  0.038 \\
H. Gille &  0.101 &  0.183 &  0.011 &  0.165 &  \textbf{0.535} &  0.347 &  0.023 &  0.082 &  0.144 &  \textbf{0.424} &  \textbf{0.366} &  0.339 &  0.263 \\
A. Lichtwitz  &  \textbf{0.193} &  0.113 &  0.206 &  0.066 &  0.053 &  0.025 &  0.020 &  0.006 &  \textbf{0.211} &  0.083 &  0.092 &  0.096 &  0.087 \\
D. Hioco &  0.140 &  0.162 &  0.003 & \textbf{ 0.315} &  0.110 &  \textbf{0.265} &  0.003 &  0.011 &  0.033 &  0.050 &  0.072 &  0.041 &  0.157 \\
De S\`{e}ze &  0.012 &  0.002 &  \textbf{0.762} &  0.002 &  0.005 &  0.003 &  \textbf{0.371} &  \textbf{0.106} &  0.001 &  0.004 &  0.003 &  0.003 &  0.003 \\
\bottomrule
\end{tabular}}
}
\end{table}

\subsection{Summary}
In Appendix \ref{sec:A1}, we showed how our method works out in full using illustrative networks, and then compared the results with those obtained with the $BL$ method. In this example, four authors with their publication lists of $10$ years are given. By checking the publication history of those authors, indeed we can confirm that the second and the third authors are experts in different topics. Our method was able to correctly identify the expertise of each author. However, the $BL$ method gave a result according to which the research profiles of the two authors were the same. This example and the comparison between methods thus showed that our method outperformed the $BL$ one.   

In Appendix \ref{sec:A2}, we gave an example of a handpicked paper, and provided the results obtained using our method. We showed that our method correctly assigned expertise to the most experienced author on most $MeSH$ terms. And authors would not acquire much experience in categories that they were not familiar with. The result showed that our method was able to add appropriate value to the co-authors’ expertise vectors and update them so that they could better represent the evolution of co-authors’ expertise.

Despite the lack of ground truth data to definitively validate the performance of our method, the examples in the Appendix provide some possible ways to test our method. The results showed that our method can provide a reasonable assessment of authors' expertise.

\color{black}

\bibliographystyle{spbasic}      %
\bibliography{ref}   %

\begin{thebibliography}{48}
\providecommand{\natexlab}[1]{#1}
\providecommand{\url}[1]{{#1}}
\providecommand{\urlprefix}{URL }
\expandafter\ifx\csname urlstyle\endcsname\relax
  \providecommand{\doi}[1]{DOI~\discretionary{}{}{}#1}\else
  \providecommand{\doi}{DOI~\discretionary{}{}{}\begingroup
  \urlstyle{rm}\Url}\fi
\providecommand{\eprint}[2][]{\url{#2}}

\bibitem[{AlShebli et~al.(2018)AlShebli, Rahwan, and
  Woon}]{alshebli2018preeminence}
AlShebli BK, Rahwan T, Woon WL (2018) The preeminence of ethnic diversity in
  scientific collaboration. Nature Communications 9(1):5163

\bibitem[{Balog et~al.(2007)Balog, De~Rijke et~al.}]{balog2007determining}
Balog K, De~Rijke M, et~al. (2007) Determining expert profiles (with an
  application to expert finding). In: IJCAI, vol~7, pp 2657--2662

\bibitem[{Balog et~al.(2012)Balog, Fang, de~Rijke, Serdyukov, Si
  et~al.}]{balog2012expertise}
Balog K, Fang Y, de~Rijke M, Serdyukov P, Si L, et~al. (2012) Expertise
  retrieval. Foundations and Trends{\textregistered} in Information Retrieval
  6(2--3):127--256

\bibitem[{Bao and Zhai(2017)}]{bao2017dynamic}
Bao P, Zhai C (2017) Dynamic credit allocation in scientific literature.
  Scientometrics 112(1):595--606

\bibitem[{Begum et~al.(2016)Begum, Rajesh, and Vinnarasi}]{begummeta}
Begum SSF, Rajesh A, Vinnarasi M (2016) Meta path based top-k similarity join
  in heterogeneous information networks. arXiv:161009769 [csSI]

\bibitem[{Berendsen et~al.(2013)Berendsen, De~Rijke, Balog, Bogers, and Van
  Den~Bosch}]{berendsen2013assessment}
Berendsen R, De~Rijke M, Balog K, Bogers T, Van Den~Bosch A (2013) On the
  assessment of expertise profiles. Journal of the American Society for
  Information Science and Technology 64(10):2024--2044

\bibitem[{Blei et~al.(2003)Blei, Ng, and Jordan}]{blei2003latent}
Blei DM, Ng AY, Jordan MI (2003) Latent dirichlet allocation. Journal of
  Machine Learning research 3(Jan):993--1022

\bibitem[{Duan et~al.(2012)Duan, Li, Li, Lu, and Wen}]{duan2012mei}
Duan D, Li Y, Li R, Lu Z, Wen A (2012) Mei: Mutual enhanced infinite
  community--topic model for analyzing text-augmented social networks. The
  Computer Journal 56(3):336--354

\bibitem[{Fortunato et~al.(2018)Fortunato, Bergstrom, B{\"o}rner, Evans,
  Helbing, Milojevi{\'c}, Petersen, Radicchi, Sinatra, Uzzi
  et~al.}]{fortunato2018science}
Fortunato S, Bergstrom CT, B{\"o}rner K, Evans JA, Helbing D, Milojevi{\'c} S,
  Petersen AM, Radicchi F, Sinatra R, Uzzi B, et~al. (2018) Science of science.
  Science 359(6379):eaao0185

\bibitem[{Foulkes and Neylon(1996)}]{foulkes1996redefining}
Foulkes W, Neylon N (1996) Redefining authorship. relative contribution should
  be given after each author's name. BMJ: British Medical Journal
  312(7043):1423

\bibitem[{Gerlach et~al.(2018)Gerlach, Peixoto, and
  Altmann}]{gerlach2018network}
Gerlach M, Peixoto TP, Altmann EG (2018) A network approach to topic models.
  Science Advances 4(7):eaaq1360

\bibitem[{Hertzum and Pejtersen(2000)}]{hertzum2000information}
Hertzum M, Pejtersen AM (2000) The information-seeking practices of engineers:
  searching for documents as well as for people. Information Processing \&
  Management 36(5):761--778

\bibitem[{Hirsch(2005)}]{hirsch2005index}
Hirsch JE (2005) An index to quantify an individual's scientific research
  output. Proceedings of the National Academy of Sciences of the United States
  of America 102(46):16569--16572

\bibitem[{Hirsch(2007)}]{hirsch2007does}
Hirsch JE (2007) Does the h index have predictive power? Proceedings of the
  National Academy of Sciences of the United States of America
  104(49):19193--19198

\bibitem[{Hofmann et~al.(2010)Hofmann, Balog, Bogers, and
  De~Rijke}]{hofmann2010contextual}
Hofmann K, Balog K, Bogers T, De~Rijke M (2010) Contextual factors for finding
  similar experts. Journal of the American Society for Information Science and
  Technology 61(5):994--1014

\bibitem[{Koopman et~al.(2010)Koopman, Powers, Wang, and Wei}]{koopman2010give}
Koopman R, Powers W, Wang Z, Wei SJ (2010) Give credit where credit is due:
  Tracing value added in global production chains. Tech. rep., National Bureau
  of Economic Research

\bibitem[{Lawrence(2007)}]{lawrence2007mismeasurement}
Lawrence PA (2007) The mismeasurement of science. Current Biology
  17(15):R583--R585

\bibitem[{Lin et~al.(2006)Lin, Lyu, and King}]{lin2006pagesim}
Lin Z, Lyu MR, King I (2006) Pagesim: a novel link-based measure of web page
  aimilarity. In: Proceedings of the 15th International Conference on World
  Wide Web, ACM, pp 1019--1020

\bibitem[{Meng et~al.(2014)Meng, Shi, Li, Zhang, and Wu}]{meng2014relevance}
Meng X, Shi C, Li Y, Zhang L, Wu B (2014) Relevance measure in large-scale
  heterogeneous networks. In: Asia-Pacific Web Conference, Springer, pp
  636--643

\bibitem[{Newman(2004)}]{newman2004coauthorship}
Newman ME (2004) Coauthorship networks and patterns of scientific
  collaboration. Proceedings of the National Academy of Sciences of the United
  States of America 101(suppl 1):5200--5205

\bibitem[{Nguyen and Bai(2010)}]{nguyen2010cosine}
Nguyen HV, Bai L (2010) Cosine similarity metric learning for face
  verification. In: Asian Conference on Computer Vision, Springer, pp 709--720

\bibitem[{Pirotte et~al.(2007)Pirotte, Renders, Saerens
  et~al.}]{pirotte2007random}
Pirotte A, Renders JM, Saerens M, et~al. (2007) Random-walk computation of
  similarities between nodes of a graph with application to collaborative
  recommendation. IEEE Transactions on Knowledge \& Data Engineering 3:355--369

\bibitem[{Ramage et~al.(2009)Ramage, Rafferty, and Manning}]{ramage2009random}
Ramage D, Rafferty AN, Manning CD (2009) Random walks for text semantic
  similarity. In: Proceedings of the 2009 workshop on graph-based methods for
  natural language processing, Association for Computational Linguistics, pp
  23--31

\bibitem[{Rosen-Zvi et~al.(2004)Rosen-Zvi, Griffiths, Steyvers, and
  Smyth}]{rosen2004author}
Rosen-Zvi M, Griffiths T, Steyvers M, Smyth P (2004) The author-topic model for
  authors and documents. In: Proceedings of the 20th Conference on Uncertainty
  in Artificial Intelligence, AUAI Press, pp 487--494

\bibitem[{Rybak et~al.(2014)Rybak, Balog, and
  N{\o}rv{\aa}g}]{rybak2014temporal}
Rybak J, Balog K, N{\o}rv{\aa}g K (2014) Temporal expertise profiling. In:
  European Conference on Information Retrieval, Springer, pp 540--546

\bibitem[{Serdyukov et~al.(2011)Serdyukov, Taylor, Vinay, Richardson, and
  White}]{serdyukov2011automatic}
Serdyukov P, Taylor M, Vinay V, Richardson M, White RW (2011) Automatic people
  tagging for expertise profiling in the enterprise. In: European Conference on
  Information Retrieval, Springer, pp 399--410

\bibitem[{Shen and Barab{\'a}si(2014)}]{shen2014collective}
Shen HW, Barab{\'a}si AL (2014) Collective credit allocation in science.
  Proceedings of the National Academy of Sciences 111(34):12325--12330

\bibitem[{Shi et~al.(2012)Shi, Kong, Yu, Xie, and Wu}]{shi2012relevance}
Shi C, Kong X, Yu PS, Xie S, Wu B (2012) Relevance search in heterogeneous
  networks. In: Proceedings of the 15th International Conference on Extending
  Database Technology, ACM, pp 180--191

\bibitem[{Shi et~al.(2014)Shi, Kong, Huang, Philip, and Wu}]{shi2014hetesim}
Shi C, Kong X, Huang Y, Philip SY, Wu B (2014) Hetesim: A general framework for
  relevance measure in heterogeneous networks. IEEE Transactions on Knowledge
  and Data Engineering 26(10):2479--2492

\bibitem[{Silva et~al.(2018)Silva, Ribeiro, and Silva}]{silva2018hierarchical}
Silva J, Ribeiro P, Silva F (2018) Hierarchical expert profiling using
  heterogeneous information networks. In: International Conference on Discovery
  Science, Springer, pp 344--360

\bibitem[{Smalheiser and Torvik(2009)}]{smalheiser2009author}
Smalheiser NR, Torvik VI (2009) Author name disambiguation. Annual Review of
  Information Science and Technology 43(1):1--43

\bibitem[{Stallings et~al.(2013)Stallings, Vance, Yang, Vannier, Liang, Pang,
  Dai, Ye, and Wang}]{stallings2013determining}
Stallings J, Vance E, Yang J, Vannier MW, Liang J, Pang L, Dai L, Ye I, Wang G
  (2013) Determining scientific impact using a collaboration index. Proceedings
  of the National Academy of Sciences of the United States of America
  110(24):9680--9685

\bibitem[{Sun et~al.(2009{\natexlab{a}})Sun, Han, Zhao, Yin, Cheng, and
  Wu}]{sun2009rankclus}
Sun Y, Han J, Zhao P, Yin Z, Cheng H, Wu T (2009{\natexlab{a}}) Rankclus:
  integrating clustering with ranking for heterogeneous information network
  analysis. In: Proceedings of the 12th International Conference on Extending
  Database Technology: Advances in Database Technology, ACM, pp 565--576

\bibitem[{Sun et~al.(2009{\natexlab{b}})Sun, Yu, and Han}]{sun2009ranking}
Sun Y, Yu Y, Han J (2009{\natexlab{b}}) Ranking-based clustering of
  heterogeneous information networks with star network schema. In: Proceedings
  of the 15th ACM SIGKDD International Conference on Knowledge Discovery and
  Data Mining, ACM, pp 797--806

\bibitem[{Sun et~al.(2011)Sun, Han, Yan, Yu, and Wu}]{sun2011pathsim}
Sun Y, Han J, Yan X, Yu PS, Wu T (2011) Pathsim: Meta path-based top-k
  similarity search in heterogeneous information networks. Proceedings of the
  VLDB Endowment 4(11):992--1003

\bibitem[{Tang(2016)}]{tang2016aminer}
Tang J (2016) Aminer: Toward understanding big scholar data. In: Proceedings of
  the ninth ACM International Conference on Web Search and Data Mining, ACM, pp
  467--467

\bibitem[{Tang et~al.(2008)Tang, Jin, and Zhang}]{tang2008topic}
Tang J, Jin R, Zhang J (2008) A topic modeling approach and its integration
  into the random walk framework for academic search. In: 2008 Eighth IEEE
  International Conference on Data Mining, IEEE, pp 1055--1060

\bibitem[{Torvik and Smalheiser(2009)}]{torvik2009author}
Torvik VI, Smalheiser NR (2009) Author name disambiguation in medline. ACM
  Transactions on Knowledge Discovery from Data (TKDD) 3(3):11

\bibitem[{Tsatsaronis et~al.(2011)Tsatsaronis, Varlamis, Torge, Reimann,
  N{\o}rv{\aa}g, Schroeder, and Zschunke}]{tsatsaronis2011become}
Tsatsaronis G, Varlamis I, Torge S, Reimann M, N{\o}rv{\aa}g K, Schroeder M,
  Zschunke M (2011) How to become a group leader? or modeling author types
  based on graph mining. In: International Conference on Theory and Practice of
  Digital Libraries, Springer, pp 15--26

\bibitem[{Tscharntke et~al.(2007)Tscharntke, Hochberg, Rand, Resh, and
  Krauss}]{tscharntke2007author}
Tscharntke T, Hochberg ME, Rand TA, Resh VH, Krauss J (2007) Author sequence
  and credit for contributions in multiauthored publications. PLoS Biology
  5(1):e18

\bibitem[{Van~Gysel et~al.(2016)Van~Gysel, de~Rijke, and
  Worring}]{van2016unsupervised}
Van~Gysel C, de~Rijke M, Worring M (2016) Unsupervised, efficient and semantic
  expertise retrieval. In: Proceedings of the 25th International Conference on
  World Wide Web, International World Wide Web Conferences Steering Committee,
  pp 1069--1079

\bibitem[{Van~Rijnsoever and Hessels(2011)}]{van2011factors}
Van~Rijnsoever FJ, Hessels LK (2011) Factors associated with disciplinary and
  interdisciplinary research collaboration. Research Policy 40(3):463--472

\bibitem[{Wang et~al.(2015)Wang, Liu, Desai, Danilevsky, and
  Han}]{wang2015constructing}
Wang C, Liu J, Desai N, Danilevsky M, Han J (2015) Constructing topical
  hierarchies in heterogeneous information networks. Knowledge and Information
  Systems 44(3):529--558

\bibitem[{Wang et~al.(2016)Wang, Sun, Song, Han, Song, Wang, and
  Zhang}]{wang2016relsim}
Wang C, Sun Y, Song Y, Han J, Song Y, Wang L, Zhang M (2016) Relsim: relation
  similarity search in schema-rich heterogeneous information networks. In:
  Proceedings of the 2016 SIAM International Conference on Data Mining, SIAM,
  pp 621--629

\bibitem[{Wang et~al.(2012)Wang, Hu, Tu, and He}]{wang2012author}
Wang J, Hu X, Tu X, He T (2012) Author-conference topic-connection model for
  academic network search. In: Proceedings of the 21st ACM International
  Conference on Information and Knowledge Management, ACM, pp 2179--2183

\bibitem[{Xiong et~al.(2015)Xiong, Zhu, and Philip}]{xiong2015top}
Xiong Y, Zhu Y, Philip SY (2015) Top-k similarity join in heterogeneous
  information networks. IEEE Transactions on Knowledge and Data Engineering
  27(6):1710--1723

\bibitem[{Xu et~al.(2014)Xu, Shi, Qiao, Zhu, Jung, Lee, and
  Choi}]{xu2014author}
Xu S, Shi Q, Qiao X, Zhu L, Jung H, Lee S, Choi SP (2014) Author-topic over
  time (atot): a dynamic users’ interest model. In: Mobile, ubiquitous, and
  intelligent computing, Springer, pp 239--245

\bibitem[{Yao et~al.(2014)Yao, Mak et~al.}]{yao2014pathsimext}
Yao K, Mak HF, et~al. (2014) Pathsimext: revisiting pathsim in heterogeneous
  information networks. In: International Conference on Web-Age Information
  Management, Springer, pp 38--42

\end{thebibliography}

\end{document}